\pdfoutput=1
\documentclass[a4paper,fleqn,usenatbib]{mnras}

\usepackage{hyperref}
\usepackage{bookmark}

\usepackage[T1]{fontenc}
\usepackage{ae,aecompl}
\usepackage{times}

\usepackage{subfloat}
\usepackage{graphicx}	
\usepackage{amsmath}	
\usepackage{amssymb}	

\title[SDSS-IV MaNGA: gradients and environment]{SDSS-IV MaNGA: stellar population gradients as a function of galaxy environment}

\author[Goddard et al.]{D.~Goddard$^{1}$\thanks{E-mail: daniel.goddard@port.ac.uk},
D.~Thomas$^{1}$,
C.~Maraston$^{1}$,
K.~Westfall$^{1}$,
J.~Etherington$^{1}$,
R.~Riffel$^{2, 3}$, 
\newauthor{N.~D.~Mallmann$^{2,3}$, Z.~Zheng$^{4}$, M.~Argudo-Fern{\'a}ndez$^{5}$, M.~Bershady$^{6}$, K.~Bundy$^{7}$,}
\newauthor{N.~Drory$^{8}$, D.~Law$^{9}$, R.~Yan$^{10}$, D.~Wake$^{6,11}$, A.~Weijmans$^{12}$, D.~Bizyaev$^{13,14}$,}
\newauthor{J.~Brownstein$^{15}$, R. R.~Lane$^{16}$, R.~Maiolino$^{17,18}$, K.~Masters$^{1}$, M.~Merrifield$^{19}$,}
\newauthor{C.~Nitschelm$^{20}$, K.~Pan$^{13}$, A.~Roman-Lopes$^{21}$, T.~Storchi-Bergmann$^{2,3}$}\\
Affiliations are listed at the end of the paper}

\date{Accepted XXX. Received YYY; in original form ZZZ}

\pubyear{2016}

\begin{document}
\bibliographystyle{mnras}
\label{firstpage}
\pagerange{\pageref{firstpage}--\pageref{lastpage}}
\maketitle


\begin{abstract}
We study the internal radial gradients of stellar population properties within $1.5\;R_{\rm e}$ and analyse the impact of galaxy environment. We use a representative sample of 721 galaxies with masses ranging between $10^{9}\;M_{\odot}$ to $10^{11.5}\;M_{\odot}$ from the SDSS-IV survey MaNGA. We split this sample by morphology into early-type and late-type galaxies. Using the full spectral fitting code FIREFLY, we derive the light and mass-weighted stellar population properties age and metallicity, and calculate the gradients of these properties. We use three independent methods to quantify galaxy environment, namely the $N^{th}$ nearest neighbour, the tidal strength parameter $Q$ and distinguish between central and satellite galaxies. In our analysis, we find that early-type galaxies generally exhibit shallow light-weighted age gradients in agreement with the literature and mass-weighted median age gradients tend to be slightly positive. Late-type galaxies, instead, have negative light-weighted age gradients. We detect negative metallicity gradients in both early and late-type galaxies that correlate with galaxy mass, with the gradients being steeper and the correlation with mass being stronger in late-types. We find, however, that stellar population gradients, for both morphological classifications, have no significant correlation with galaxy environment for all three characterisations of environment. Our results suggest that galaxy mass is the main driver of stellar population gradients in both early and late-type galaxies, and any environmental dependence, if present at all, must be very subtle.
\end{abstract}



\section{Introduction}
The current paradigm for the evolution of the universe involves a cosmological constant $\Lambda$ associated with dark energy and cold dark matter (CDM). The $\Lambda$CDM model \citep{white1978,white1985}, postulates that following a hot big bang, a period of exponential growth, known as `inflation' occurred \citep{guth1981}. This expansion produced the homogeneity and isotropy of the universe. Cold dark matter particles collapsed under their own self gravity to form dark matter halos and these halos then merged, deepening the gravitational potential. Accretion of baryonic matter into these halos produced the primordial seeds of galaxy formation. The evolution of these `proto-galaxies' through cosmic time produced the structures that are observed today.
\\
\\
During this evolution, a galaxy experiences a wide range of interactions that are dependent on its location relative to other galaxies in the Universe, or known more commonly as the galaxies' environment. Dense environments, such as clusters, expose galaxies to interactions such as tidal stripping \citep{read2006}, galaxy harassment \citep{farouki1981} or even strangulation of gas from neighbours \citep{larson1980}. Galaxies that reside in under denser regions, such as voids however, remain largely untouched, accreting gas from the intergalactic medium. This diverse range of evolutionary processes should affect the galaxies properties in different ways. Yet, until the discovery of the morphology-density relation \citep{oemler1974, dressler1980}, the importance of environment on galaxy evolution was poorly understood. Since then, studies on the impact of environment on galaxy properties have become an active area of research. \\
\\
Modern spectroscopic galaxy surveys such as the Sloan Digital Sky Survey (SDSS, \cite{york2000}) and the Two-degree Field Galaxy Redshift Survey (2dFGRS, \cite{colless2003}) have contributed largely to these studies by providing a statistical sample of a million galaxies, allowing the galaxies in the local Universe to be probed in much finer detail. Studies have shown that global properties such as colour, star formation rate and stellar age \citep{hogg2004,mercurio2010,kauffmann2004,peng2010,thomas2010}, have only a mild dependence on environment. Studies have also shown that parameters related to structures, such as Sersic index and surface brightness are nearly independent of environment \citep{blanton2005,blanton2009}. This discord of results has made it difficult to establish to what extent environment is a pivotal driver in galaxy evolution. 
\\
\\
One downside of these large spectroscopic surveys is that only a small subregion of the galaxy is sampled, defined by the location of the light collecting fiber. Therefore neglecting the complex and rich internal structure of galaxies where important environmental effects, on properties such stellar population gradients, might be seen \citep{barbera2011}. In order to decipher the internal components of the galaxy and understand in detail the dependence of stellar population gradients on environment, it is necessary to use integral field spectroscopy (IFS). A number of spatially resolved measurements on local galaxies have already been made (SAURON \citep{deze2002}, DiskMass \citep{bershady2010}, ATLAS$^{3D}$ \citep{cappellari2011}, CALIFA \citep{sanchez2012}, SAMI survey \citep{allen2015}), which have provided evidence for inside-out mass assembly of galaxies \citep{perez2013} and `sub maximality' of disks \citep{bershady2011}, and probed the internal chemical composition of galaxies. There has also been efforts to map stellar content of galaxies in very high spatial resolution IFS data \citep[e.g.][]{bacon1995,mcdermid2006,ricdavies2007, brasil1,brasil2,brasil3, kaman2016}. \\
\\
Despite this there have only been a small number of IFU studies focusing on stellar population gradients in a statistical manner, which is partly down to the smaller sample sizes used in previous surveys. MaNGA (Mapping Nearby Galaxies at Apache Point, \citet{bundy2015}), which is part of the fourth generation of SDSS, aims to complement these previous surveys by offering a large statistical sample of 10,000 nearby galaxies (median redshift $z \sim 0.03$) with extensive wavelength coverage (3600$\AA$-10300$\AA$) by 2020. This large wavelength coverage is useful for breaking the age/metallicity degeneracy. Crucially for this work, MaNGA allows us to resolve galaxies spatially out to at least 1.5 effective radii (R$_{e}$) from a wide range of environments. Additionally, the survey also provides a flat distribution of galaxies in the $i$-band absolute magnitude (M$_{i}$ as a proxy for stellar mass), enabling us to robustly assess how stellar population gradients vary across different mass galaxies from different environments. A parallel MaNGA paper by \citet{zheng2016} also investigates the impact of galaxy environment on stellar population gradients using independent fitting codes, stellar population models and tracers of galaxy environment. An explanation on the complementary nature of this work and differences in methodology will be explicated throughout this text.
\\
\\
This paper is organised in the following manner; Section 2 explains details of the MaNGA survey and the numerical tools used for full spectral fitting and obtaining radial gradients. Section 3 provides a detailed introduction on the methodology used to determine galaxy environment. In Section 4, we present the results of our study, then briefly provide a discussion in Section 5 and finally describe our conclusions in Section 6. Throughout this paper, the redshifts and stellar masses quoted are taken from the Nasa Sloan Atlas catalogue (NSA1, \cite{blantoncat}). When quoting luminosities, masses and distances, we make use of a $\Lambda$CDM cosmology with $\Omega_{m} = 0.3$ and H$_{0} = 67$ km$^{-1}$ s$^{-1}$ Mpc$^{-1}$ \citep{planck2015}.

\begin{figure*}
\includegraphics[width=0.45\textwidth]{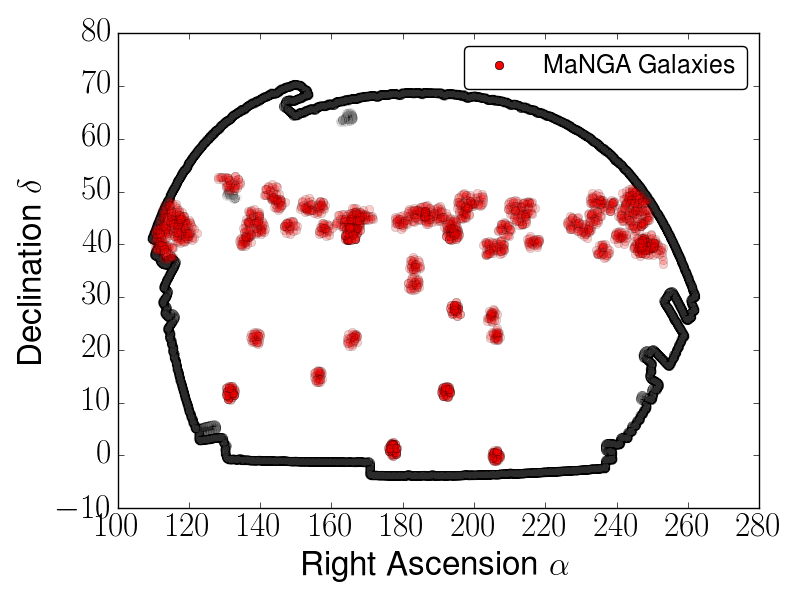}
\includegraphics[width=0.45\textwidth]{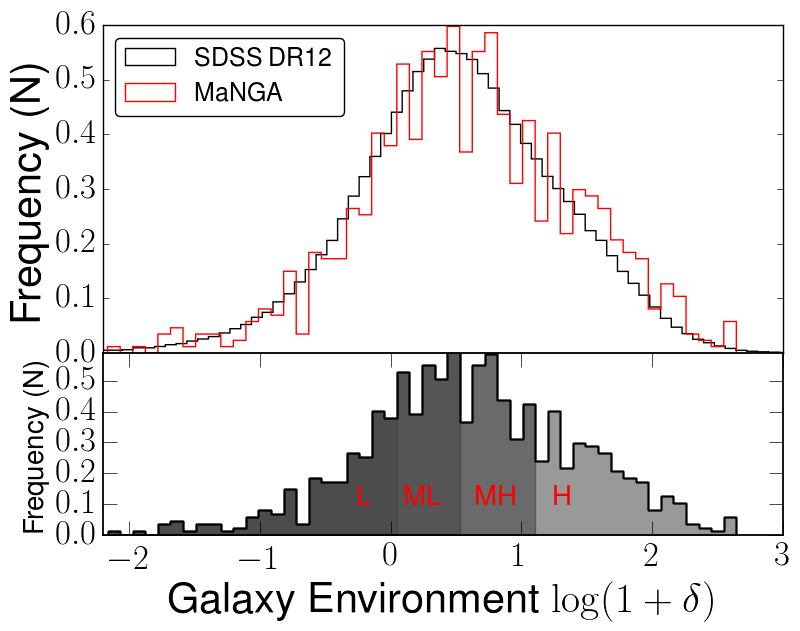}
\caption{The left hand plot shows a section of the SDSS footprint and the red circles highlight the positions of the observed MaNGA galaxies. The right hand plot shows a comparison of galaxy environments derived for the SDSS DR12 sub-sample and for the MaNGA galaxy in this paper. For the DR12 sub-sample we imposed a redshift cut of $z < 0.15$ (as this is the upper redshift limit of the MaNGA survey) and a magnitude cut in the $r$ band of $M_r= -20$. The red labels L, ML, MH and H correspond to the different environmental density percentiles.}
\label{fig:environment}
\end{figure*}

\section{Data and Stellar Population Analysis} 
In a companion paper \citep[hereafter Paper 1]{goddard2016a}, we present a comprehensive description of our data analysis and provide an assessment of the use of different spectral fitting codes and stellar population models. We also explore the correlation between stellar population gradients, galaxy mass and morphology. In this Section, we briefly highlight some of the key information and refer the reader to the paper for more details.

\subsection{The MaNGA Survey}
The MaNGA survey \citep{bundy2015} is part of the fourth generation of the Sloan Digital Sky Survey (SDSS) and aims to obtain spatially resolved spectroscopy of 10,000 nearly galaxies (median redshift $z\sim 0.03$) by 2020. MaNGA uses 5 different types of integral field unit (IFU), with sizes that range from 19 fibres ($12.5''$ diameter) to 127 fibres ($32.5''$ diameter), to optimise these observations. Fibre bundle size and galaxy redshift are selected such that the fibre bundle provides the desired radial coverage (see Wake et al. in prep for further details on sample selection and bundle size optimisation and \citet{law2015} for observing strategy). In this work, we selected an original sample of 806 galaxies from the MaNGA data release MPL4 (equivalent to the public release SDSS DR13, \url{www.sdss.org/dr13}), that were observed during the first year of operation. The observational data was reduced using the MaNGA data-reduction-pipeline (DRP,  \citet{law2016}) and then analysed using the MaNGA data analysis pipeline (DAP, Westfall et al, in prep). To classify galaxies by morphology, we used Galaxy Zoo \citep{lintott2011}. In this work we split the galaxies into two subsets, namely `Early-type' galaxies (Elliptical/Lenticular) and `Late-type' galaxies (Spiral/Irregular). Galaxies with an 80$\%$ majority vote for a specific morphological type from the Galaxy Zoo were selected for this analysis. Galaxies which did not fulfil this criterion were visually inspected and classified by the authors.

\subsection{Full Spectral Fitting}
The spectral fitting code FIREFLY (see \citet{wilkinson2015} for more details) and the models of \cite{maraston2011} are used to derive stellar population properties from MaNGA Data Analysis Pipeline (DAP) Voronoi binned spectra with $S/N > 5$. This is different to the work of \citet{zheng2016}, where the full spectral fitting code STARLIGHT \citep{starlight2005} and stellar population models of \citet{bruzual2003}, with a \citet{chabrier2003} initial mass function (IMF), are used. A full comparison on the choice of fitting codes and models can be found in Paper 1.
\\
\\
FIREFLY uses a $\chi^2$ minimisation technique\footnote{Calculated as $\chi^2= \sum_{\lambda} \frac{(O(\lambda)-M(\lambda))^2}{E(\lambda)^2}$, where $O(\lambda)$ is the observed SED, $M(\lambda)$ is the model spectrum and $E(\lambda)$ is the error.} that given an input spectral energy distribution (SED), returns a set of typically 100-1000 model fits. These initial fits are then checked to see whether their $\chi^2$ values can be improved by adding a different Simple Stellar Population (SSP) component with luminosity equal to the first one. This process is then iterated until the $\chi^2$ is minimised and the solution cannot be improved by a statistically significant amount, which is governed by the Bayesian Information Criterion (BIC, \cite{liddle2007}). Prior to fitting the model templates to the data, FIREFLY takes into account galactic and interstellar reddening of the spectra. Foreground Milky Way reddening is accounted for by using the foreground dust maps of \cite{schlegel1998} and the extinction curve from \cite{fitzpatrick1999}. The dust attenuation of each source is determined in the following way. The model templates and data are preprocessed using a `High Pass Filter (HPF)'. The HPF uses an analytic function across all wavelengths to rectify the continuum before deriving the stellar population parameters, allowing the removal of large scale features (continuum shape and dust extinction). \\
\\
FIREFLY requires two additional inputs provided by the DAP; measurements of the stellar velocity dispersion $\sigma$ and fits to the strong nebular lines. Stellar velocity dispersion is needed to effectively remove the influence of the stellar kinematics on the stellar population fit, and the determination of this is done using the Penalized Pixel-Fitting (pPXF) method of \cite{emsellem}. The MaNGA DAP also fits individual Gaussians to the strong nebular emission lines after subtracting the best-fit stellar-continuum model from pPXF. The best-fitting parameters for all the fitted lines [\ion{O}{II}], [\ion{O}{III}], [\ion{O}{I}], H$\alpha$,  H$\beta$, [\ion{N}{II}], and [\ion{S}{II}] are used to construct a model, emission-line only spectrum for each binned spectrum.  These models are subtracted from the binned spectra to produce emission-free spectra for analysis using FIREFLY.

\subsection{Radial Gradients}
\begin{figure*}
\centering
\raisebox{4.5mm}{\includegraphics[width=0.18\textwidth, height=0.18\textwidth]{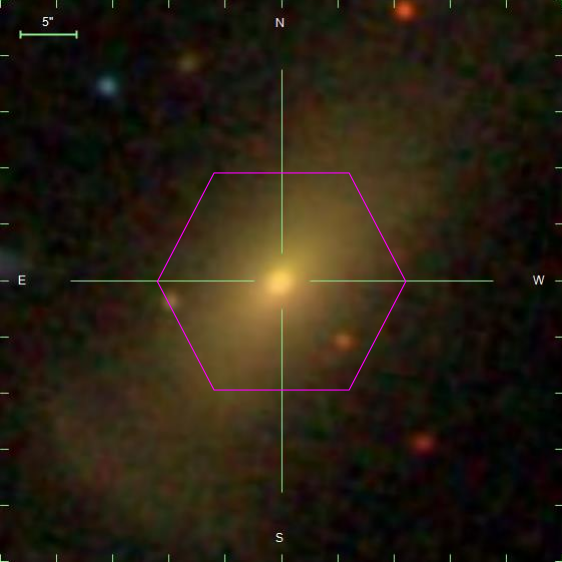}}
\hspace{0.1cm}
\includegraphics[width=0.26\textwidth, height=0.20\textwidth]{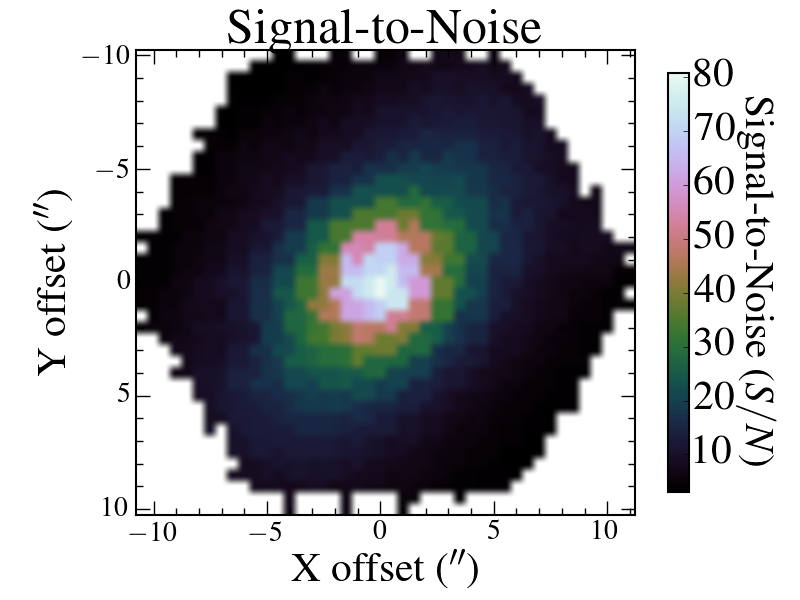}
\includegraphics[width=0.26\textwidth, height=0.20\textwidth]{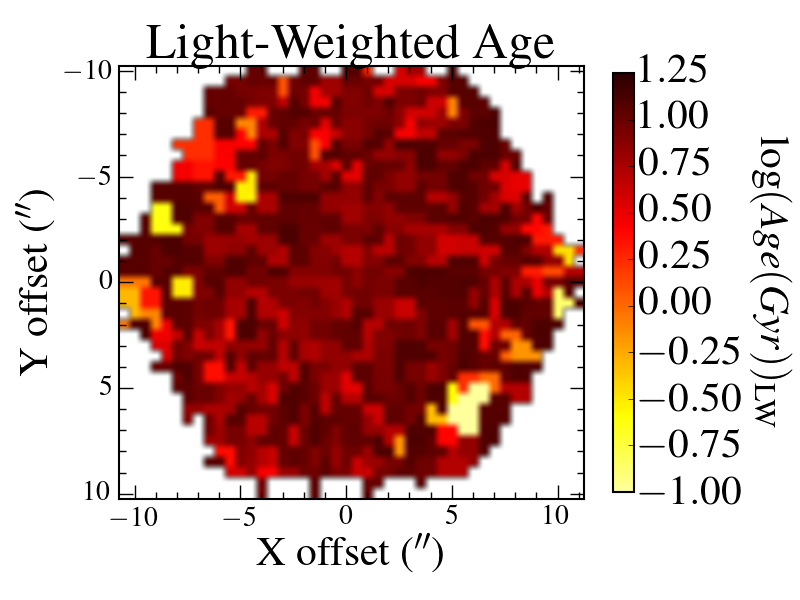}
\includegraphics[width=0.26\textwidth, height=0.20\textwidth]{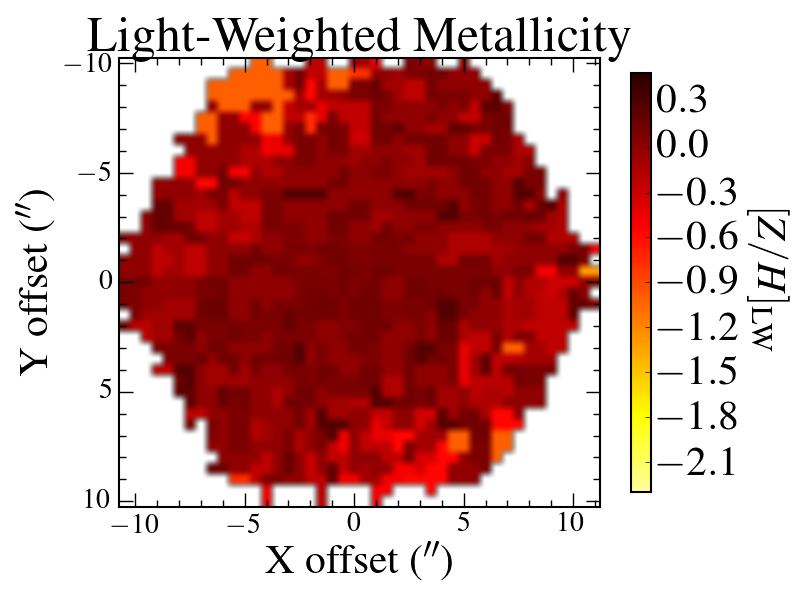}
\includegraphics[width=0.35\textwidth, height=0.30\textwidth]{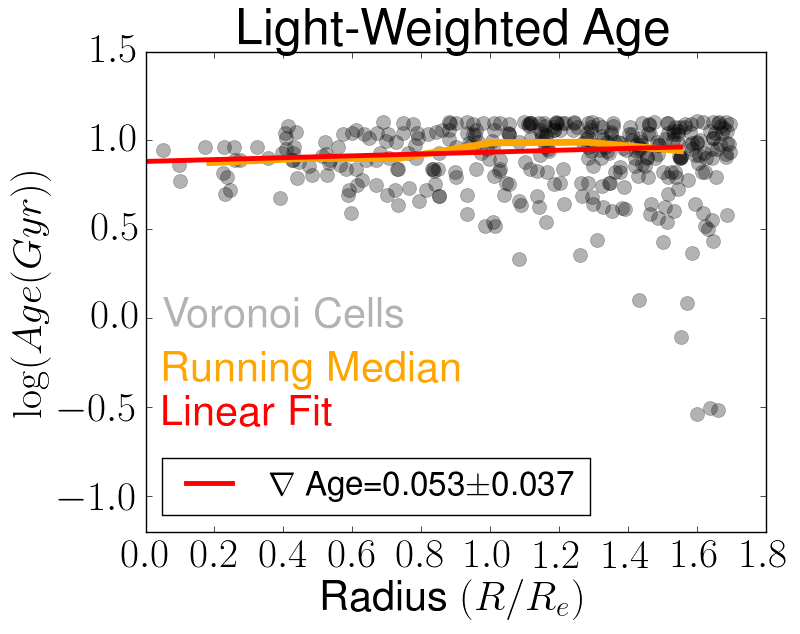}
\hspace{0.15cm}
\includegraphics[width=0.35\textwidth, height=0.30\textwidth]{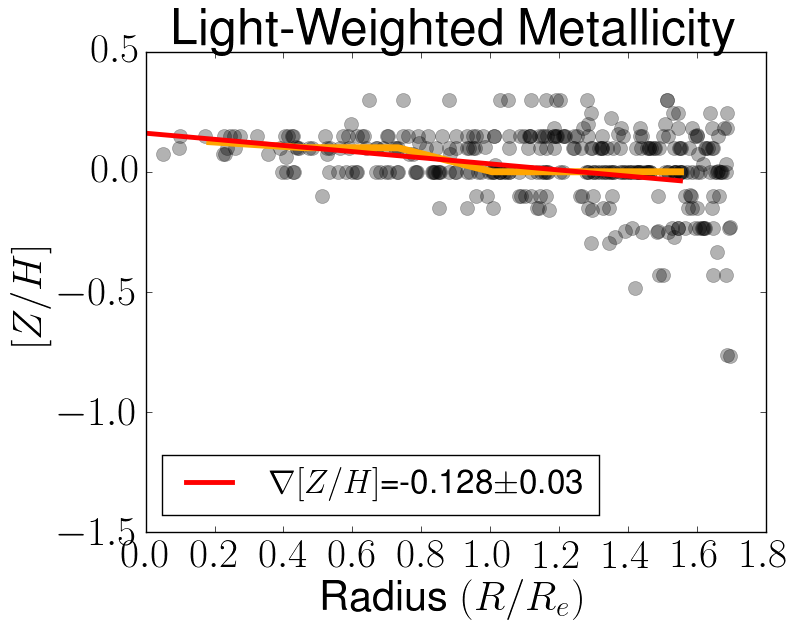}
\caption{An example early-type galaxy from the MaNGA survey (MaNGA ID 1-114998) that has been observed with the 61 fibre IFU. The top row (from left to right) shows the SDSS image of the galaxy, the corresponding signal-to-noise map, and the light-weighted age and metallicity maps derived from FIREFLY, respectively. The bottom row shows the radial profiles of light-weighted age and metallicity for the galaxy, where the grey circles represent individual Voronoi cells from the DAP data cube, the orange line shows the running median and the red line shows the least squares fit. The gradient value and corresponding error is quoted in the legend.}
\label{fig:sp_maps}
\end{figure*}

The effective radius $R_{\rm e}$, position angle and ellipticity of each galaxy is measured from Sloan Digital Sky Survey photometry by performing a one component, two-dimensional Se\'rsic fit in the $r$-band \citep{blantoncat}. The on-sky position (relative to the galaxy centre) of each Voronoi cell is then used to calculate semi-major axis coordinates, which we then use to define a radius $R$ of the cell. We define the radial gradient of a stellar population property $\theta$ (e.g. $\log(Age(Gyr))$, $[Z/H]$) in units of dex/$R_{e}$ as:
\begin{equation}
\nabla \theta= \mathrm{d}\theta/\mathrm{d}R,
\end{equation}
where $R$ is the radius in units of effective radius $R_{\rm e}$. The gradient is measured using least squares linear regression (see Figure~\ref{fig:sp_maps}) . Errors on the gradients are calculated using a Monte Carlo bootstrap resampling method \citep{recipes}. 

\subsection{Final Sample}
Due to the complex geometry of the SDSS footprint (which consists of an array of parabolic strips), some MaNGA galaxies that reside close to the footprint edge had to be excluded from the analysis because an accurate measure of environment was not possible (see left panel of Figure~\ref{fig:environment}). Furthermore, a number of galaxies that were in the final morphologically classified sample had to be neglected from the final analysis due to having unreliable velocity dispersion estimates from the DAP. This led to the exclusion of 85 galaxies (33 early-type galaxies and 52 late-type galaxies spanning a range of environments and masses) from our original sample of 806 galaxies, leaving 505 early-type galaxies and 216 late-type galaxies (70\% and 30\% of the sample respectively). 

\section{Galaxy Environment}
A galaxy's environment is often expressed as the density field in which it resides. To quantify galaxy environment, a plethora of different indicators can be used. This can range from fixed aperture methods, which involves choosing a circle of radius $r$ around the galaxy in question and counting how many galaxies fall inside this circle giving a number density, to more complex methods taking into account redshift space distortions \citep{cooper2005,schawinski2007}, and tidal tensor prescriptions based on the Hessian of the gravitational potential \citep{eardley2015}. These methods probe different environmental scales, so it is essential to choose the appropriate method that explores the desired range of the study. 
\\
\\
In this work, we measure galaxy environment using $N^{th}$ nearest neighbour local number density, gravitational tidal strength, and classifying between central and satellite galaxies. In \citet{zheng2016}, environment is described using a galaxies location in the Large Scale Structure (LSS); being categorised into either a cluster, filament, sheet or void environment (see \citet{hahn2007, wang2009,wang2012}). A comparison of the \citet{zheng2016} study, and the results presented here, will be provided in the Discussion.

\subsection{Local Density}
\label{sec:envdef}
\begin{figure}
\includegraphics[width=0.49\textwidth]{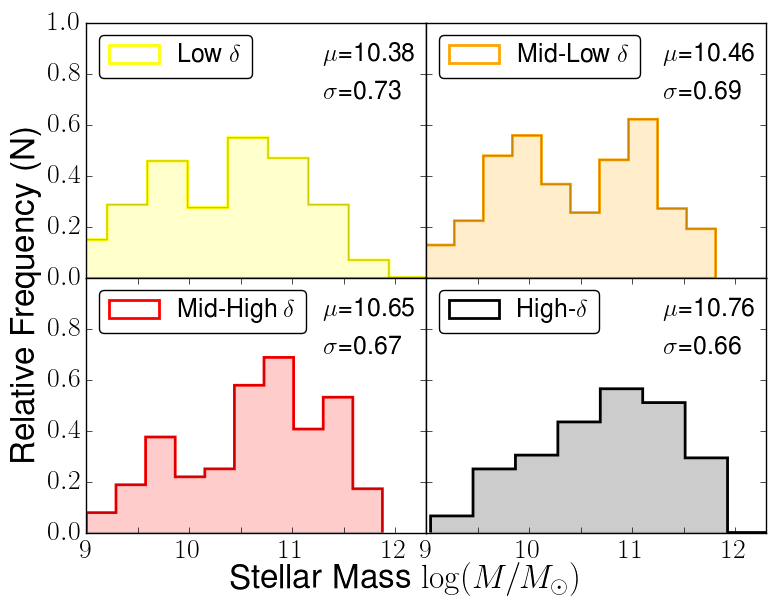}
\caption{Distributions of stellar mass ($\log(M/M_{\odot})$) for the four different environmental bins that make up the sample used in this work. Galaxy masses are drawn from the Nasa Sloan Atlas catalogue (NSA1). The mean $\mu$ and 1-$\sigma$ value of each distribution are quoted in the corresponding panel.}
\label{fig:final_sample}
\end{figure}

In this work, we look at local galaxy environment, which is well determined using $N^{th}$ Nearest neighbour methods; see \citet{muldrew2012} for a review. This method requires choosing a number $N$ of neighbours, calculating the distance to the $N^{th}$ neighbour and constructing a volume with this radius. Dividing $N$ by this volume gives the number density. Dense environments are obtained when the $N^{th}$ nearest neighbour is close to the target galaxy. The redshift range for neighbouring galaxies is $\pm \Delta$ zc = 1000 km/s. We select $N=5$ and utilise an algorithm developed in \citet{etherington2015}. It was shown in \citet{baldry2007} that the best estimate of local environment was an average of $N$= 4 and 5, hence we chose a value of $N$ close to this to obtain robust measurements.
\\
\\  
A local overdensity $\delta$ is defined as:
\begin{equation}
\delta=\frac{\rho_i -\rho_m}{\rho_m}
\end{equation}
where $\rho_i$ is the number density described using $N^{th}$ neighbour and $\rho_m$ is the mean density of galaxies within a redshift window centred on the target galaxy utilising all of the available area. A galaxies environment is then given by:
\begin{equation}
\log(1+\delta)
\end{equation}
From these measurements, we construct the distribution of environments for the MaNGA galaxy sample and compare this to the distribution of environments calculated for a magnitude and redshift matched sample of SDSS DR12 galaxies \citep{dr12}. This is to ensure that we were not biasing our measurements and only sampling MaNGA galaxies from particular environmental densities. The right hand panel of Figure~\ref{fig:environment} shows this distribution of environments for the MaNGA galaxy sample compared to the environments of the SDSS DR12 sample. This demonstrates that the environmental densities of the MaNGA sample used here are representative of the environmental density distribution derived from a much larger, statistically complete sample.\\
\\
We split the MaNGA environment distribution into quartiles to define four different environmental densities (see bottom right panel of Figure~\ref{fig:environment}). Galaxies were then assigned to one of these groups. 
\begin{itemize}
\item{$\delta$ < 25th percentile  = {\it Low $\delta$}}
\item{25th percentile < $\delta$ < 50th percentile = {\it Mid-Low $\delta$}} 
\item{50th percentile < $\delta$ < 75th percentile = {\it Mid-High $\delta$}}
\item{$\delta$ > 75th percentile = {\it High $\delta$}}
\end{itemize}
The number of galaxies in each environmental bin for our final analysis is 180, 178, 182 and 181 respectively. The distribution of galaxy masses that make up these bins can be seen in Figure~\ref{fig:final_sample}. It can be seen that each bin of environmental density samples the full mass range and recovers well the mass-density relation, where the most massive galaxies live in the densest environments \citep{baldry2007}.

\subsection{Mass-Dependent Environmental Measure}
\begin{figure}
\includegraphics[width=0.49\textwidth]{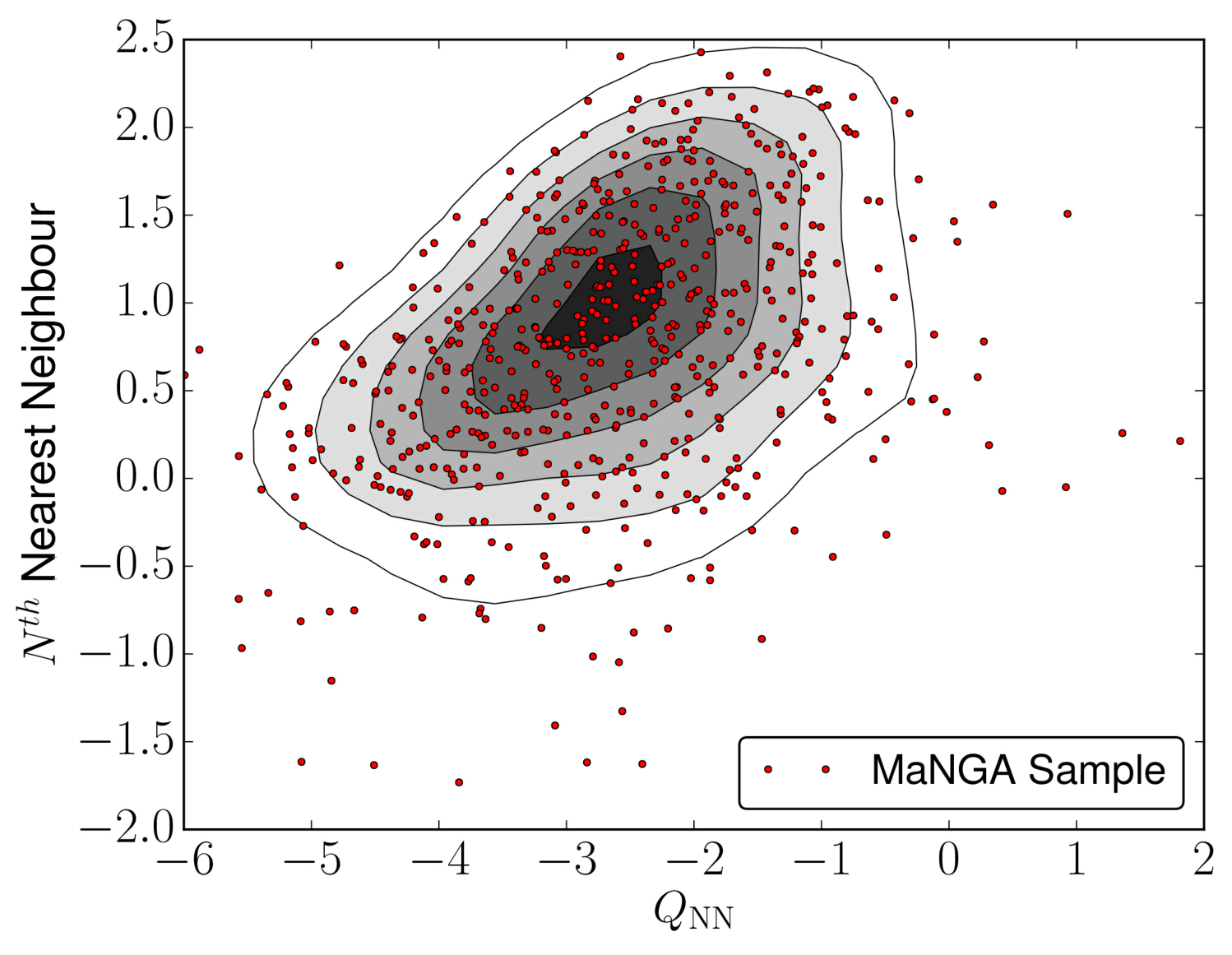}
\caption{Figure showing the comparison between the environments calculated using the $N^{th}$ nearest neighbour method and the $Q$ parameter. The red points represent individual galaxies used in this work, the grey contours represent the density of points. The Spearman's rank correlation coefficient is 0.4.}
\label{fig:env_compare_NN}
\end{figure}

Possible biases can arise when using just one environmental measure. Consider for example, galaxies in an isolated triplet system. The local number density given by the nearest neighbour method would likely determine a low density environment as the 5th neighbour may be very far away. Yet, there may be significant gravitational interaction caused by the nearby galaxies which could lead to environmental effects. As a cross reference, and to ensure the results presented in this paper are robust, we repeated our analysis using a mass-dependent measure known as the tidal strength estimator, $Q$. The tidal strength estimator quantifies the strength of gravitational interaction that nearby neighbouring galaxies inflict on a central galaxy with respect to its internal binding forces \citep{maria1,maria2,maria3}. For one neighbour, $Q_{ip}$ is given by:
\begin{equation}
Q_{ip} = \frac{F_{\text{Tidal}}}{F_{\text{Binding}}} \propto \frac{M_{i}}{M_{p}} \left(\frac{D_{p}}{R_{ip}}\right)^{3}
\end{equation}
where $M_{i}$ is the mass of the neighbouring galaxy, $M_{p}$ is the mass of the primary galaxy, $D_{p}$ is the apparent diameter of the galaxy estimated by an isophote containing 90\% of the total $r$-band flux of the galaxy and $R_{ip}$ is the projected distance between the neighbour and primary galaxy. Assuming a linear mass-luminosity relation \citep{bell2003, bell2006} the stellar mass is proportional to the $r$-band flux at a fixed distance, with $m_{r}$ = -2.5$\log(\mathrm{flux}_{r})$. The formula for one neighbour can be written:
\begin{equation}
Q_{ip} = 0.4\left(m_{r}^{p} - m_{r}^{i} \right) + 3\log \left(\frac{D_{p}}{R_{ip}}\right) 
\end{equation}
where $m_{r}^{p}$ and $m_{r}^{i}$ are the apparent magnitudes in the $r$-band of the primary galaxy and the neighbour respectively. The tidal parameter $Q$ for $n$ galaxies is then defined as the dimensionless quantity of the gravitational interaction strength created by all the neighbours in the field:
\begin{equation}
Q = \log \left(\sum_{i = 1}^{n} Q_{ip} \right)
\end{equation}
A low value of $Q$ implies that the primary galaxy is well isolated from external influences. The Spearman's rank correlation coefficient between the environments calculated from the $N^{th}$ nearest neighbour method and the $Q$ parameter is 0.4 (see Figure~\ref{fig:env_compare_NN}). 

\subsection{Central and Satellite Galaxies}
Most galaxies in the Universe are situated in many body systems. This can range from dense clusters of thousands of galaxies, to galaxy pairs. The central galaxies in clusters tend to be the most luminous and most massive galaxies in the Universe and reside at the potential minimum of the dark matter halo. These galaxies also seem to be drawn from a different luminosity function compared to most other bright elliptical galaxies \citep{bern2001}, thus hinting at a different evolutionary process. Satellite galaxies, which are galaxies moving relative to the potential minimum (having fallen into the larger halo), are also thought to have unique evolutionary signatures. Their star formation is thought to be rapidly quenched when gas is removed due to ram pressure stripping. Therefore, it is interesting to consider how stellar population gradients in central and satellite galaxies change as a function of local environment. In order to separate the MaNGA galaxy sample used in this work into central/satellite galaxies, we use the halo-based group finder developed by \citet{yang2007}. 
\\
\\
In \citet{yang2007}, all galaxies from the SDSS with $z < 0.20$ and an $r$-band magnitude brighter than 18 mags were selected and a halo based group finder was used to identify the location of galaxies within different dark matter haloes. Once these haloes had been identified, the most luminous galaxies were defined as central galaxies and the others were defined as satellite galaxies. We then cross-matched the MaNGA galaxy sample used in this work to this catalogue. We classified 478 central galaxies and 243 satellite galaxies and use our $N^{th}$ nearest neighbour measurements of environment to investigate whether stellar population gradients are different in central and satellite galaxies and whether there are possible dependencies on local environmental density. The satellite fraction of $\sim$33$\%$ used in this work is an appropriate representation of the local galaxy population, as it is similar to the fraction obtained in the larger MaNGA parent sample ($\sim$31$\%$) and to the fraction calculated at $z$ $\sim 0.03$ in the complete \citet{yang2007} catalogue ($\sim$30$\%$).

\section{Results} 
In Paper 1, we find that early-type galaxies generally exhibit shallow light-weighted age gradients ($\nabla \log(Age(Gyr))_{\mathrm{LW}}$ $\sim -0.004$ dex/$R_{e}$) and slightly positive mass-weighted age gradients ($\nabla \log(Age(Gyr))_{\mathrm{MW}}$ $\sim 0.092$ dex/$R_{e}$). Light and mass-weighted metallicity gradients tend to be negative ($\nabla [Z/H]_{\mathrm{LW}}$ $\sim -0.12$ dex/$R_{e}$, $\nabla [Z/H]_{\mathrm{MW}}$ $\sim -0.05$ dex/$R_{e}$). These values agree well with previous literature, such as \citet{mehlert2003} ($\nabla \log(Age(Gyr))_{\mathrm{LW}}$ $\sim 0$ dex, $\nabla [Z/H]_{\mathrm{LW}}$ $\sim -0.16$ dex), \citet{rawle2010} ($\nabla \log(Age(Gyr))_{\mathrm{LW}}$ $\sim -0.02$ dex$^{-1}$, $\nabla [Z/H]_{\mathrm{LW}}$ $\sim -0.13$ dex$^{-1}$), \citet{spolaor2009} ($\nabla [Z/H]_{\mathrm{LW}}$ $\sim -0.16$ dex/$R_{e}$), and modern cosmological simulations \citep{hirschmann2015}. However, our light-weighted metallicity gradients are shallower than what is found by \citet{kuntschner2010} ($\nabla [Z/H]_{\mathrm{LW}}$ $= -0.28 \pm 0.12$ dex/$R_{e}$). There are a number of possible reasons for this difference in gradient value. Firstly, the choice of stellar population models and stellar library is important when deriving gradients. It was shown in Paper 1 (and can be seen in \citet{gonz2015}), that the use of different models can lead to offsets in the derived gradients by $0.1-0.3$ dex. Secondly, depending on the spatial resolution of the data, beam smearing can flatten out the inferred radial gradient. SAURON data is used in \citet{kuntschner2010}, which has much higher spatial resolution than MaNGA data. However, the effect of beam smearing was investigated in Paper 1 and we found no significant impact on our gradients. Lastly, the radial range over which the gradient is calculated can also have a significant effect on the gradient, as it was shown in \citet{gonz2015}, that different gradients can be found in the inner and outer regions of a galaxy. A comprehensive discussion of the derived metallicity gradients from the literature is discussed in Paper 1, and the median literature metallicity gradient was found to be $\mu = -0.20$, with a spread $\sigma=0.11$. Our result, and that of \citet{kuntschner2010}, sit reasonably well within this range.
\\
\\
For Late-type galaxies, we find negative light-weighted age gradients ($\nabla \log(Age(Gyr))_{\mathrm{LW}}$ $\sim -0.11$ dex/$R_{e}$) and flat mass-weighted age gradients ($\nabla \log(Age(Gyr))_{\mathrm{MW}}$ $\sim 0.01$ dex/$R_{e}$). Both light and mass-weighted metallicity gradients are found to be negative ($\nabla [Z/H]_{\mathrm{LW}}$ $\sim -0.07$ dex/$R_{e}$, $\nabla [Z/H]_{\mathrm{MW}}$ $\sim -0.10$ dex/$R_{e}$), similar to what was found in the CALIFA survey \citep{sanchez2014,gonz2015} and consistent with the inside-out formation of disc galaxies.
\\
\\
In Paper 1, we also investigated the relationship between stellar population gradients and stellar mass by fitting linear relationships in the gradient-mass plane. We found that no correlation exists between age gradients and mass for both early and late-type galaxies. However, there is a correlation between the negative metallicity gradients and mass, where the gradients become steeper with increasing galaxy mass, agreeing with what was found in \citet{gonz2015}. In this Section, we break these results down further and investigate the relationship between the stellar population gradients of both early and late-type galaxies with galaxy environment, as described by three independent environment measures.

\subsection{Local Density}
\begin{figure*}
\includegraphics[width=0.41\textwidth]{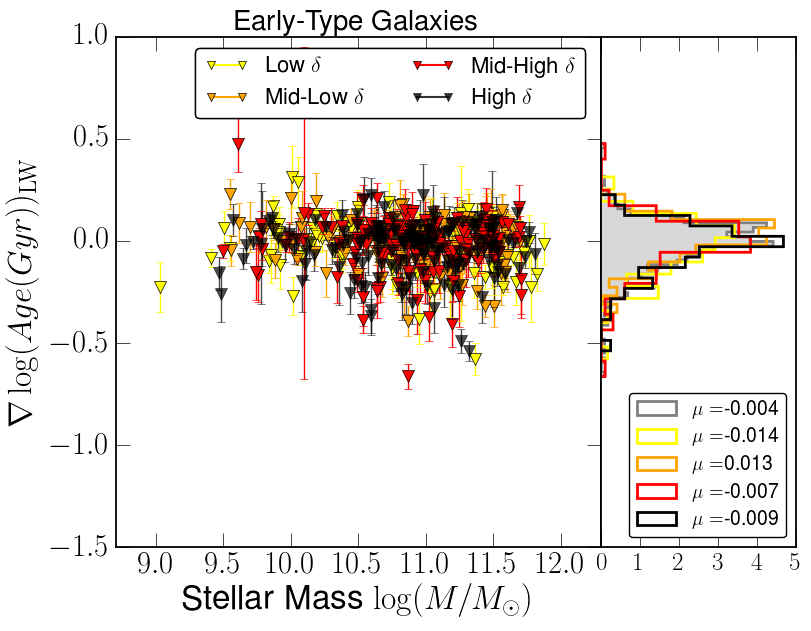}
\includegraphics[width=0.41\textwidth]{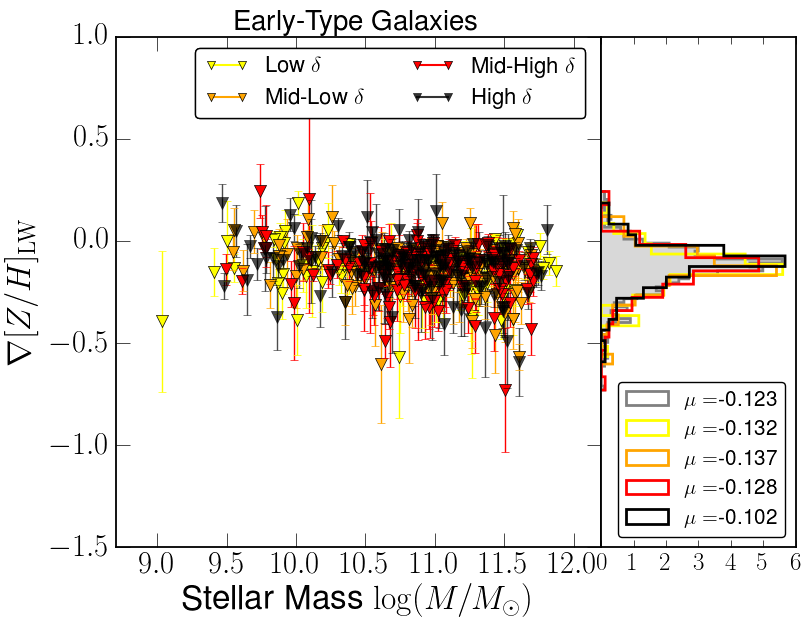}
\includegraphics[width=0.41\textwidth]{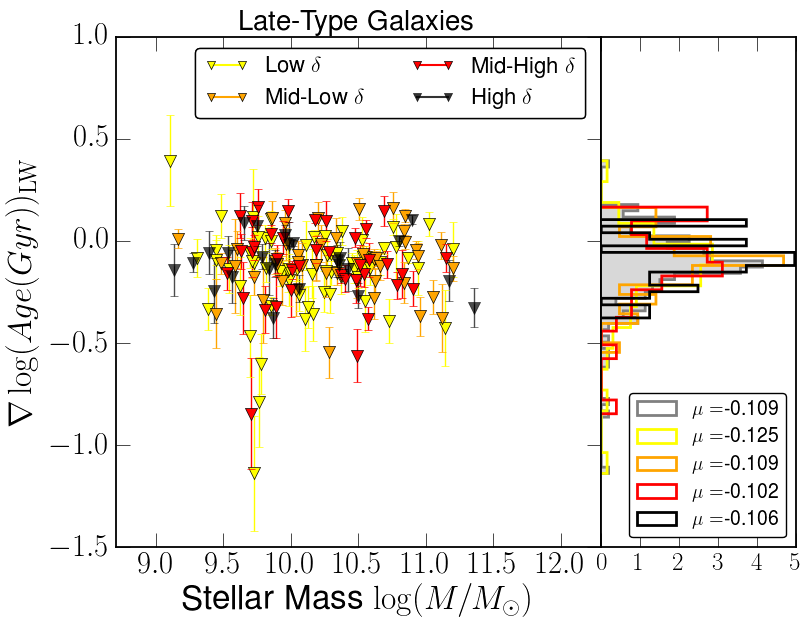}
\includegraphics[width=0.41\textwidth]{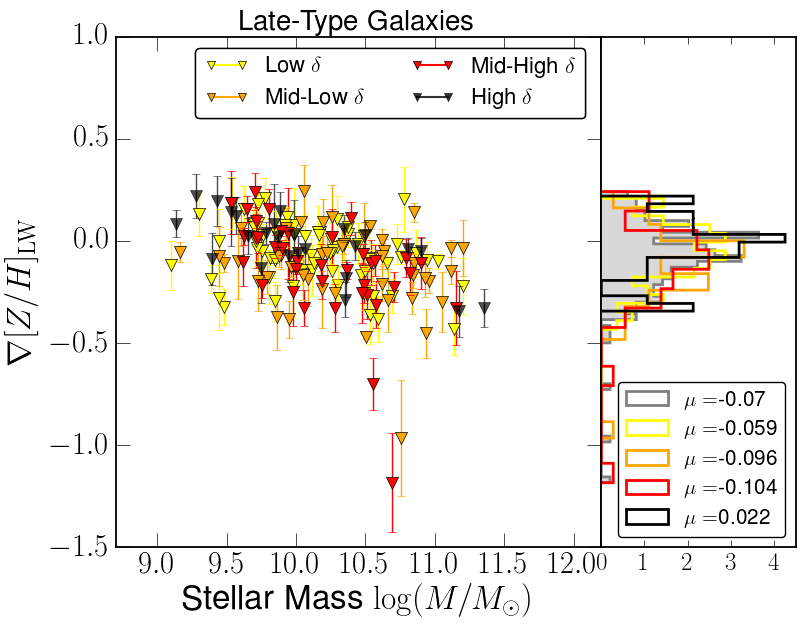}
\includegraphics[width=0.41\textwidth]{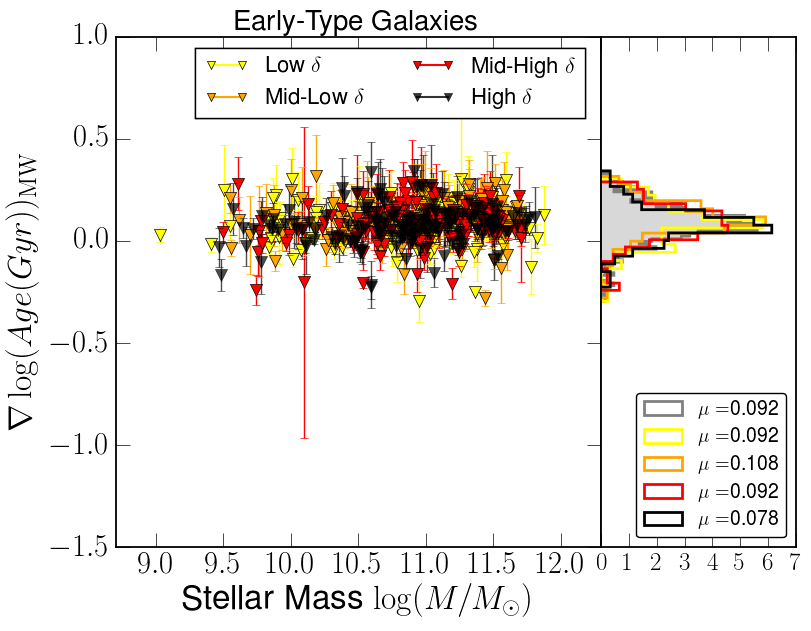}
\includegraphics[width=0.41\textwidth]{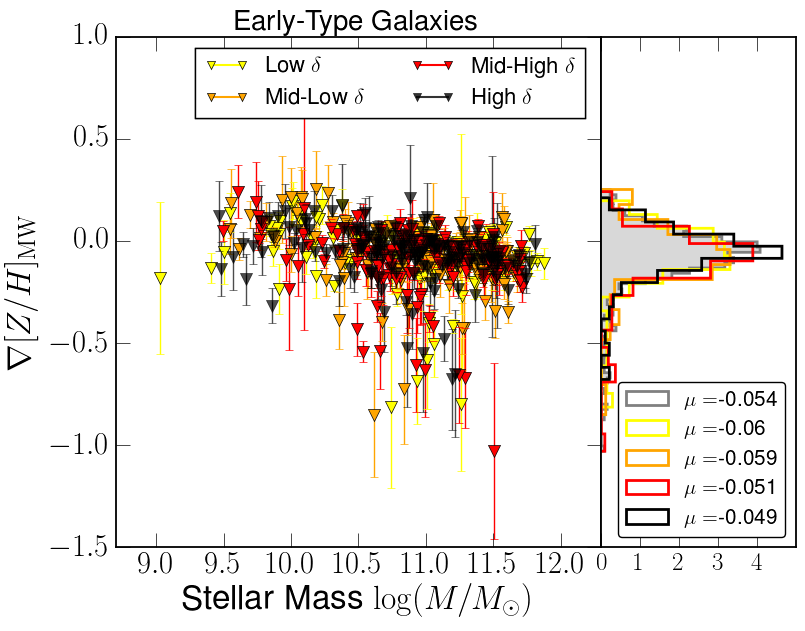}
\includegraphics[width=0.41\textwidth]{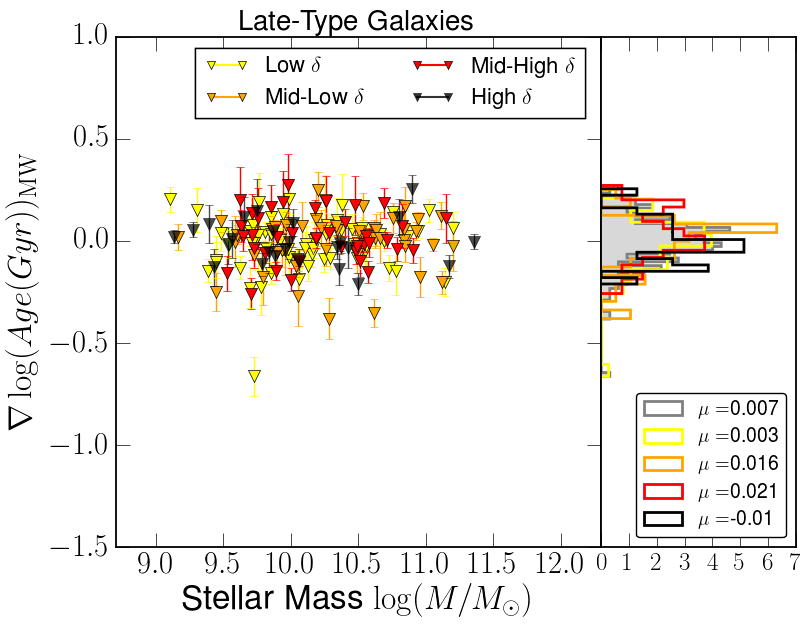}
\includegraphics[width=0.41\textwidth]{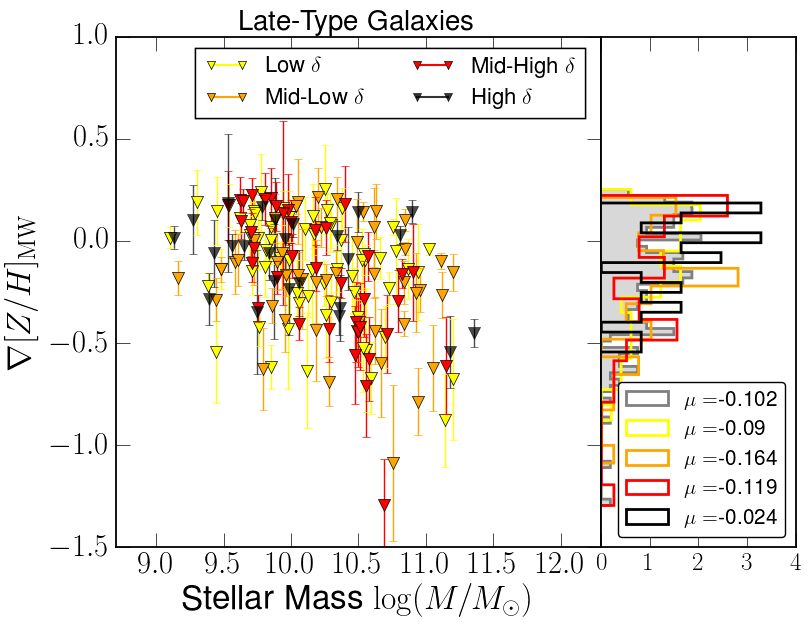}
\caption{Light and mass-weighted stellar population gradients in age (left-hand panels) and metallicity (right-hand panels) as a function of galaxy mass for different local environmental densities. Plots with axis label subscripted $\mathrm{ET}$ are for early-type galaxies and $\mathrm{LT}$ represents late-type galaxies. The different marker colours indicate the four environmental densities described in Section \ref{sec:envdef}. The right-hand sub-panels show the distribution of the gradients for the whole sample and for the four different environment bins. The median value $\mu$ for each distribution is also quoted in the legend.}
\label{fig:early_age_grad}
\end{figure*}

\begin{table*}
\caption{Median light weighted and mass weighted gradients for both early type and late type galaxies. The gradients are split by different environmental densities. Errors correspond to the 1-$\sigma$ value from the distribution.}
\begin{tabular}{c c c c c c }
\hline\hline
Morphology & Property & Low $\delta$ & Mid-Low $\delta$ & Mid-High $\delta$ & High $\delta$ \\ [0.5ex]
 &  & $\nabla$ (dex/$R_{e}$) & $\nabla$ (dex/$R_{e}$) & $\nabla$ (dex/$R_{e}$) & $\nabla$ (dex/$R_{e}$) \\ [0.5ex]
\hline
Early Type & Mass Weighted Age &  0.092 $\pm$ 0.10 & 0.108 $\pm$ 0.08 & 0.092 $\pm$ 0.08 & 0.078 $\pm$ 0.07\\ [1ex]
 & Light Weighted Age &  -0.014 $\pm$ 0.09 & 0.013 $\pm$ 0.08 & -0.007 $\pm$ 0.09 & -0.009 $\pm$ 0.07\\ [1ex]
 & Mass Weighted $[Z/H]$ &  -0.06 $\pm$ 0.09 & -0.059 $\pm$ 0.08 & -0.051 $\pm$ 0.09 & -0.049 $\pm$ 0.07\ \\[1ex]
 & Light Weighted $[Z/H]$ &  -0.132 $\pm$ 0.09 & -0.137 $\pm$ 0.08 &  -0.128 $\pm$ 0.07 & -0.102 $\pm$ 0.07 \\ [3ex]
Late Type & Mass Weighted Age & 0.03 $\pm$ 0.12 & 0.016 $\pm$ 0.15  & 0.021 $\pm$ 0.16  &  -0.01 $\pm$ 0.20 \\[1ex]
 & Light Weighted Age &  -0.125 $\pm$ 0.12  &  -0.109 $\pm$ 0.15 & -0.102 $\pm$ 0.17 & -0.106 $\pm$ 0.20 \\ [1ex]
 & Mass Weighted $[Z/H]$ & -0.09 $\pm$ 0.12 & -0.164 $\pm$ 0.15 & -0.119 $\pm$ 0.16 & -0.024 $\pm$ 0.20 \\[1ex]
 & Light Weighted $[Z/H]$ &  -0.059 $\pm$ 0.12 &  -0.096 $\pm$ 0.14 & -0.104 $\pm$ 0.15 & -0.02 $\pm$ 0.19\\ [2ex]
\hline
\end{tabular}
\label{table:complete_table}
\end{table*}

Figure~\ref{fig:early_age_grad} shows the derived light and mass-weighted stellar population gradients as a function of stellar mass $\log(M/M_{\odot})$ for the four different environmental densities defined using the $N^{th}$ nearest neighbour method. Additionally, Table~\ref{table:complete_table} shows the corresponding median gradients with 1-$\sigma$ errors for each environment. For early-type galaxies, the light and mass-weighted stellar population gradients appear to be fairly homogenous across the different environments and are in good agreement with the gradients obtained for the whole sample. Light and mass-weighted ages, for each environmental density, fluctuate around $\sim0$ dex/$R_{e}$ and $\sim0.9$ dex/$R_{e}$, and light and mass-weighted metallicities tend to be negative, with values around $\sim -0.12$ dex/$R_{e}$ and $\sim-0.05$ dex/$R_{e}$, respectively. The story is similar for late-types, where light and mass-weighted ages are fairly consistent across the different environments, yielding median gradient values of $\sim-0.1$ dex/$R_{e}$ and $\sim0$ dex/$R_{e}$. Metallicity gradients tend to have a greater scatter, but there is no significant deviation from one environmental density to another.
\\
\\
To further test our conclusions, we conducted simple Kolmogorov-Smirnov (K-S) tests on the distributions of gradients for the different environmental densities. The K-S test allows us to check whether two distributions are drawn from the same underlying distribution. If environmental effects are noticeable, there will be a significant difference when comparing the cumulative distribution functions of the two most contrasting environmental densities. For this reason, we conducted our K-S tests on the Low-$\delta$ and High-$\delta$ distributions\footnote{In an attempt to account for the errors on the individual gradients when calculating the K-S statistic and P-Value, we used Monte-Carlo bootstrap resampling. This involved resampling the datasets and recalculating the quantities of order 500 times. This provided us with an error on the K-S statistic and P-Value which should take into account the errors on the gradients.}. Results of this analysis can be seen in Table~\ref{table:ks_table}.\\
\\
\begin{table}
\caption{Table showing the corresponding Kolmogorov-Smirnov statistic ($\mathrm{K-S}$) and P-Value ($\mathrm{P}$) for the empirical cumulative distribution function (ECDF) of the lowest and highest environmental densities. As the Kolmogorov-Smirnov test only cares about the raw gradient value, errors on the $\mathrm{K-S}$ statistic and P-Value were calculated via Monte-Carlo methods to attempt to account for the errors on the individual gradients.}
\centering
\begin{tabular}{c c c c}
\hline\hline
Morphology & Property & K-S Statistic & p-value \\ [0.5ex]
\hline
Early-Types & Light-Weighted Age & $0.01 \pm 0.04$ & $0.55 \pm 0.24$ \\ [1ex]
& Mass-Weighted Age & $0.12 \pm 0.04$ & $0.37 \pm 0.16$ \\ [2ex]
& Light-Weighted $[Z/H]$ & $0.21 \pm 0.07$ & $0.01 \pm 0.10$ \\[1ex]
& Mass-Weighted $[Z/H]$ & $0.11 \pm 0.04$ & $0.48 \pm 0.20$  \\[2ex]
Late-Types & Light-Weighted Age & $0.10 \pm 0.06$ & $0.91 \pm 0.25$  \\ [1ex]
& Mass-Weighted Age & $0.24 \pm 0.09$ & $0.25 \pm 0.25$  \\ [2ex]
& Light-Weighted $[Z/H]$ & $0.15 \pm 0.06$ & $0.78 \pm 0.27$ \\[1ex]
& Mass-Weighted $[Z/H]$ & $0.16 \pm 0.07$ & $0.69 \pm 0.27$ \\[2ex]
\hline
\end{tabular}
\label{table:ks_table}
\end{table}
Overall we see that for both early and late-type galaxies, the cumulative distributions of gradients do not differ much between the lowest and highest density environments, with P-Values ranging between 0.25 and 0.91. For light-weighted metallicity gradients in early-type galaxies however, there seems to be some difference between the two distributions with P-Value $=0.01 \pm 0.10$. Thus suggesting being drawn from different underlying distributions and evidence for some environmental dependence. The error, obtained via Monte Carlo bootstrap resampling, on this value is quite large, and therefore we cannot conclusively say that there is an environmental dependence on the light-weighted metallicity gradients of early-types. 
\\
\\
As mentioned previously, in Paper 1 we look at relationships between stellar population gradients and stellar mass by fitting linear relationships in the gradient-mass plane. We can extend this exercise here by fitting these relations to each of the different environmental densities to see if there is any environmental effect on this mass dependence. Table~\ref{table:linear_relations} shows the slopes of the relationship between stellar population gradient and galaxy mass for the various environmental density bins and galaxy types. There appears to be no significant slope for both early and late-type galaxies, suggesting that there is no dependence of these relationships on environmental density.

\begin{table*}
\caption{Slopes of the (linear) relationship between stellar population gradient and galaxy mass obtained for both early and late-type galaxies in different environmental densities.}
\begin{tabular}{c c c c c c}
\hline\hline
Morphology & Environment & Light-Weighted Age & Mass-Weighted Age & Light-Weighted $[Z/H]$ & Mass-Weighted $[Z/H]$\\ [0.5ex]
\hline
Early-Types & Low $\delta$ & $0.02 \pm 0.08$ & $0.01 \pm 0.04$ & $0.04 \pm 0.06$ & $-0.07 \pm 0.08$\\[1ex]
& Mid-Low $\delta$ & $-0.01 \pm 0.07$ & $0.02 \pm 0.05$ & $-0.08 \pm 0.09 $& $-0.09 \pm 0.06$ \\[1ex]
& Mid-High $\delta$  & $-0.06 \pm 0.09$ & $0.01 \pm 0.04$ & $-0.02 \pm 0.04$& $-0.09 \pm 0.07$\\[1ex]
& High $\delta$ & $-0.02 \pm 0.05$ & $-0.03 \pm 0.08$ & $-0.04 \pm 0.07$& $-0.07 \pm 0.10$\\[2ex]
Late-Types & Low $\delta$ & $-0.08 \pm 0.07$ & $-0.02 \pm 0.05$ & $-0.10 \pm 0.07 $& $-0.19 \pm 0.07$\\[1ex]
& Mid-Low $\delta$ & $-0.04 \pm 0.08$ & $0.03 \pm 0.03$  & $-0.08 \pm 0.10 $& $-0.10 \pm 0.10$\\[1ex]
& Mid-High $\delta$  & $-0.06\pm 0.06$ & $-0.02 \pm 0.04$ & $-0.24 \pm 0.09$& $-0.41 \pm 0.16$ \\[1ex]
& High $\delta$ & $0.04 \pm 0.06$ & $0.01 \pm 0.06$& $-0.18 \pm 0.06$ & $-0.13 \pm 0.13$\\[2ex]
\hline
\end{tabular}
\label{table:linear_relations}
\end{table*}

\subsection{Mass-Dependent Environmental Measure}
\begin{figure*}
 \textbf{\large{Early-Type Galaxies}}\par\medskip
\includegraphics[width=0.95\linewidth]{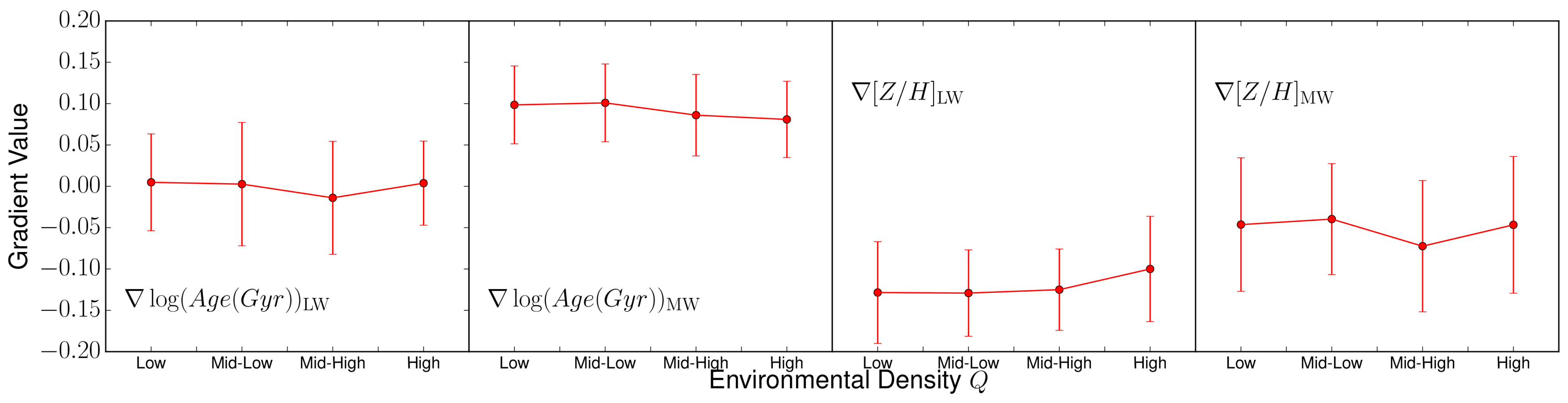}
 \textbf{\large{Late-Type Galaxies}}\par\medskip
\includegraphics[width=0.95\linewidth]{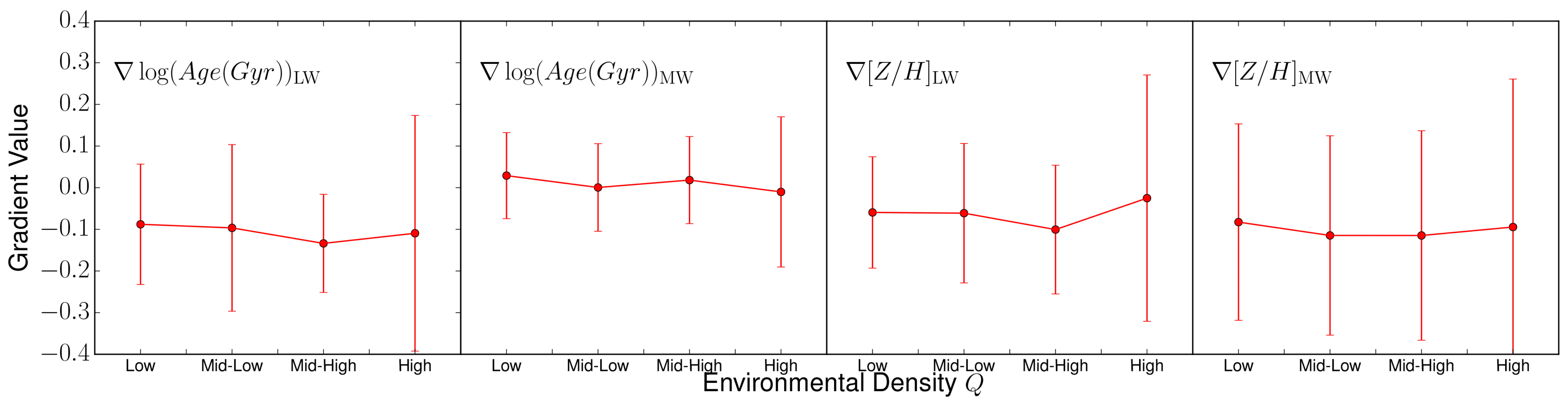}
\caption{Figure showing the median gradients obtained in different environmental densities using the $Q$ parameter. Top panels show early-type galaxies and bottom panels show late-type galaxies. From left to right, the plots show the light-weighted age, mass-weighted age, light-weighted $[Z/H]$ and mass-weighted $[Z/H]$. The error bars correspond to the standard deviation of the distribution.}
\label{fig:q_gradients}
\end{figure*}
Our analysis using the tidal strength estimator followed in exactly the same vein as before and galaxies were classified into four different environmental densities (low $Q$, mid-low $Q$, mid-high $Q$ and high $Q$). The results of this analysis is shown in Figure~\ref{fig:q_gradients}, where we plot the stellar population gradient as a function of different environmental densities. Overall, we find the the light-weighted age gradients for both early and late-type galaxies do not vary between different environments, with values of $\sim 0$ dex/$R_{e}$ and $\sim -0.1$ dex/$R_{e}$ being recovered. This is true also for the mass-weighted gradients, where gradient values in each density bin are $\sim 0.1$ dex/$R_{e}$ and $\sim 0$ dex/$R_{e}$. Light and mass-weighted metallicity gradients for both early and late-type galaxies also show no significant dependence on the environment, with median values of $\sim -0.15$ dex/$R_{e}$, $\sim -0.05$ dex/$R_{e}$, $\sim -0.05$ dex/$R_{e}$ and $\sim -0.1$ dex/$R_{e}$ being recovered in the different density bins.\\
\\
To conclude, we find no dependence of stellar population gradients on this alternative measurement of environmental density. This further strengthens the conclusion that we presented in the section above using the $N^{th}$ nearest neighbour method, that the gradients of both early and late-type galaxies are at most weakly dependent on environment.

\subsection{Central and Satellite Galaxies}
\begin{figure*}
\includegraphics[width=0.45\linewidth]{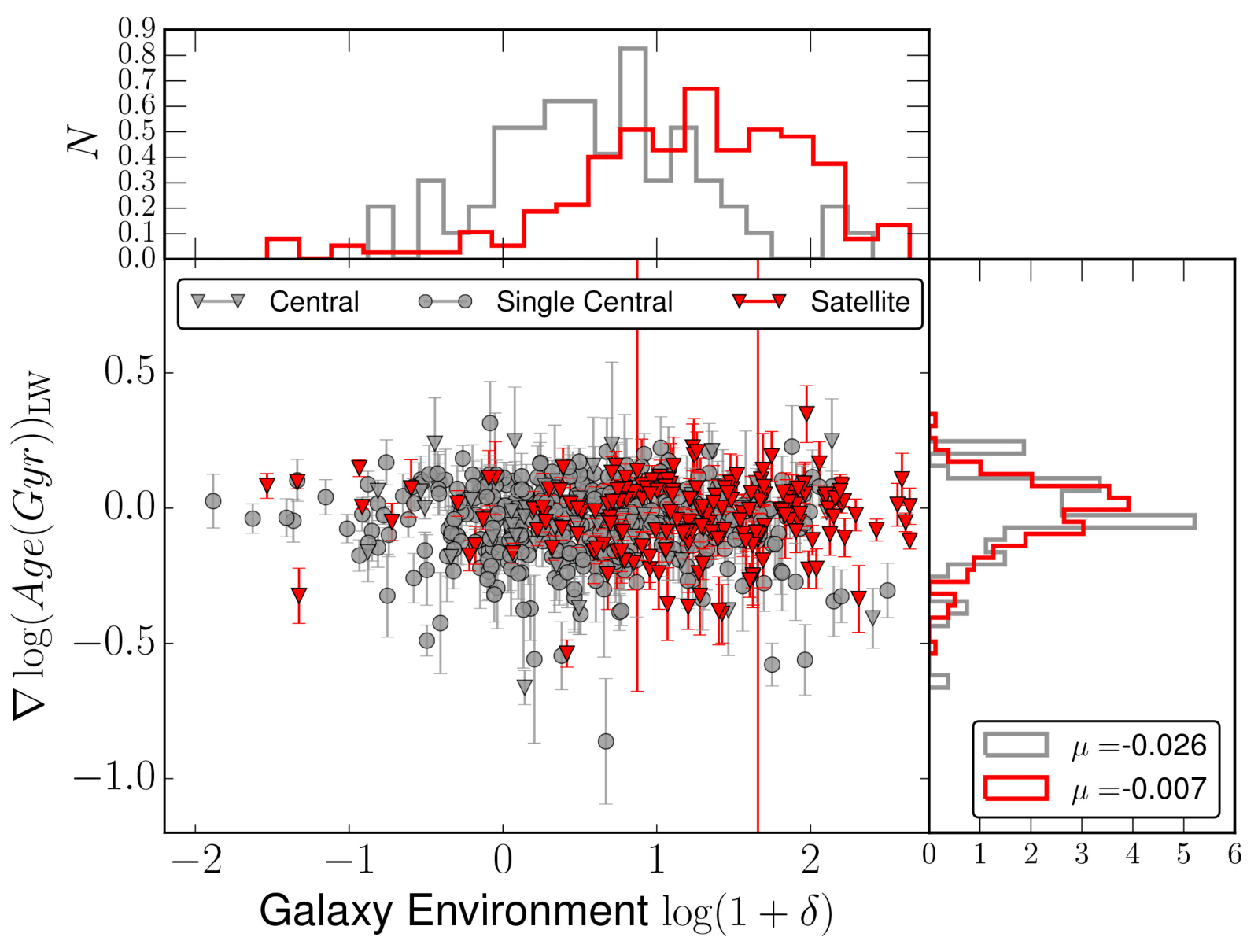}
\includegraphics[width=0.45\linewidth]{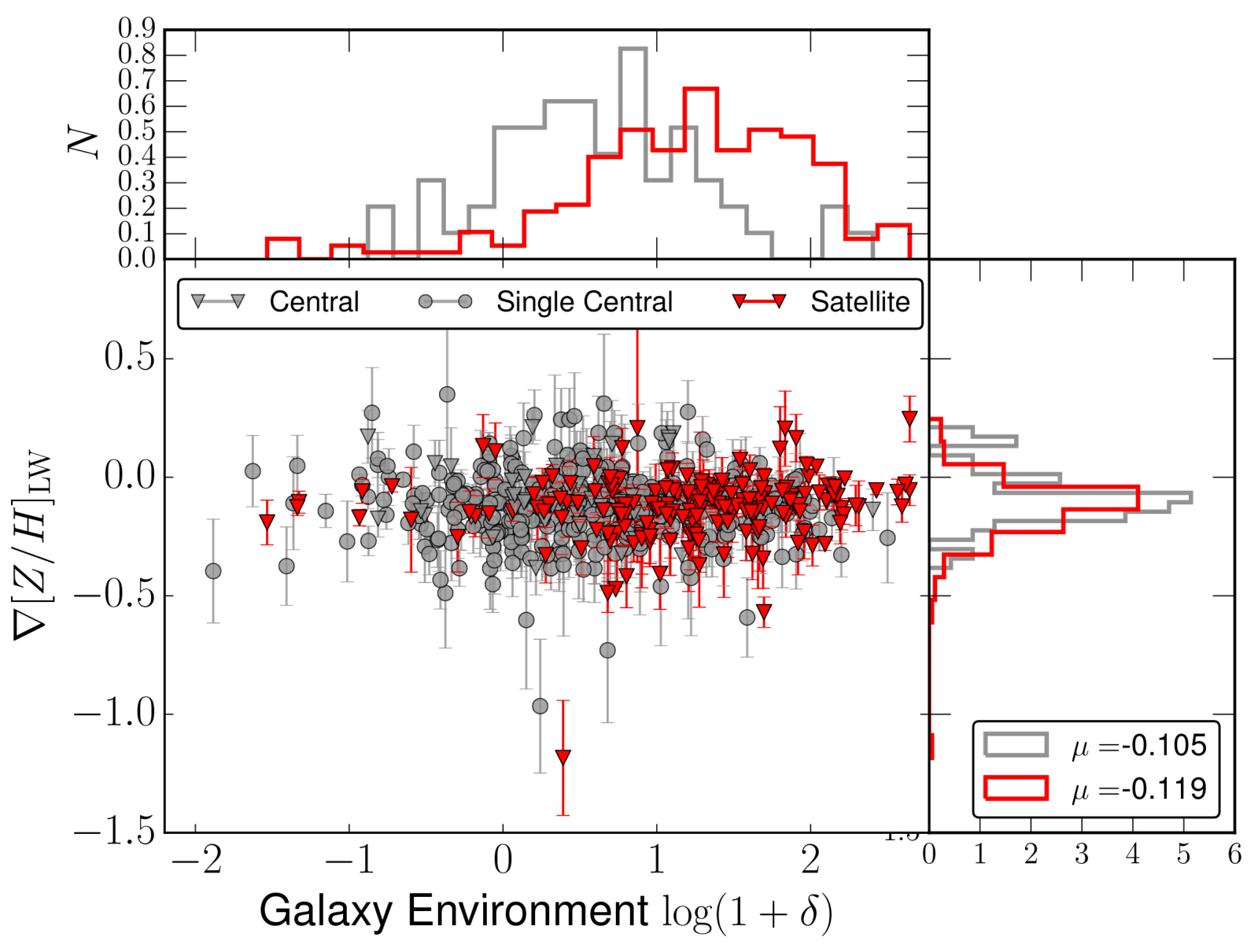}
\includegraphics[width=0.45\linewidth]{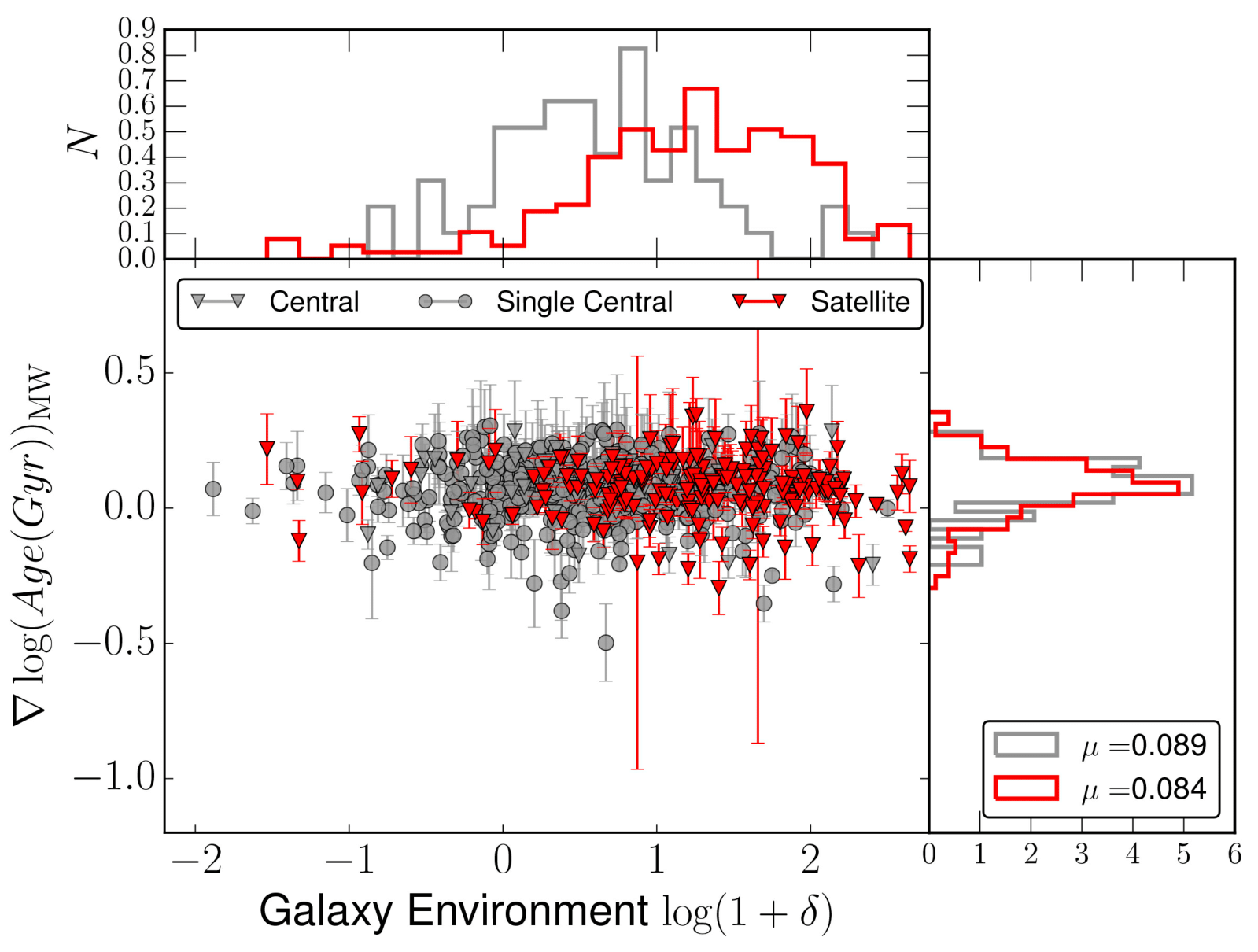}
\includegraphics[width=0.45\linewidth]{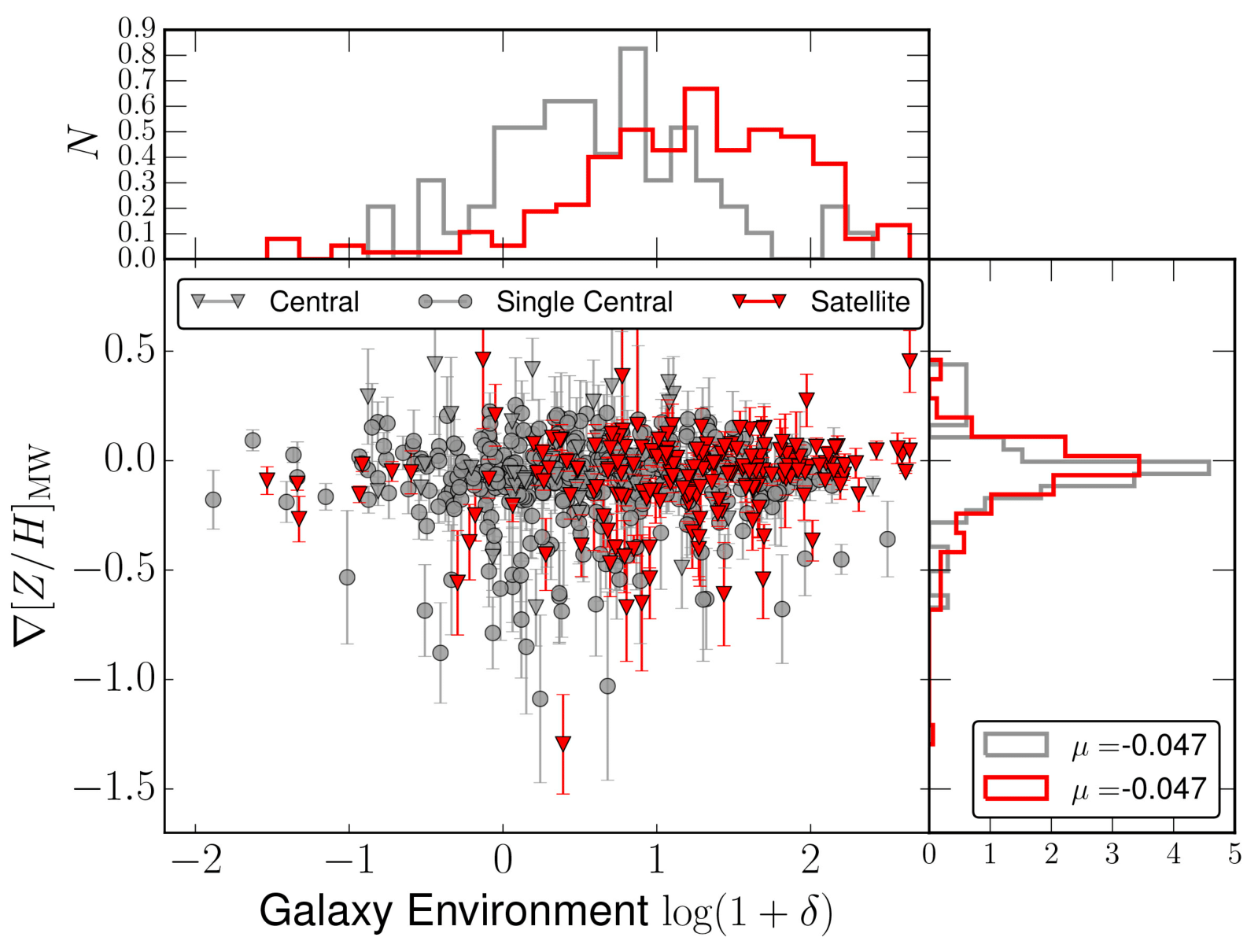}
\caption{Light (top) and mass-weighted (bottom) stellar population gradients in age and metallicity for central (grey) and satellite (red) galaxies as a function of environmental density. Central galaxies that have no satellite companions in their dark matter halo are shown by circular markers. The distributions in the top panels shows the distribution of environments for the central and satellite galaxies, the right-hand panels show the distributions of the gradients for centrals and satellites, respectively.}
\label{fig:central_satellite}
\end{figure*}
\begin{table*}
\caption{Median light and mass-weighted age/metallicity gradients obtained for the 478 central and 243 satellite galaxies from the MaNGA galaxy sample, for different environmental densities. Errors on the quantities are given by $1/\sqrt{N}$ where N is the number of galaxies in that specific bin.}
\centering
\begin{tabular}{c c c c c c }
\hline\hline
Property & Classification & Low $\delta$ &  Mid-Low $\delta$ & Mid-High $\delta$ & High $\delta$ \\ [0.5ex]
 &  & $\nabla$ (dex/$R_{e}$) & $\nabla$ (dex/$R_{e}$) & $\nabla$ (dex/$R_{e}$) & $\nabla$ (dex/$R_{e}$) \\ [0.5ex]
\hline
Light-Weighted Age & Central & $-0.01 \pm 0.05$  & $-0.04 \pm 0.06$ & $-0.03 \pm 0.06$, & $-0.01 \pm 0.05$ \\ [1ex]
& Satellite & $0.01 \pm 0.07$  & $-0.02 \pm 0.08$  &  $0.01 \pm 0.06$ & $-0.01 \pm 0.07$ \\ [2ex]
Mass-Weighted Age & Central & $0.06 \pm 0.05$  & $0.11 \pm 0.06$ &  $0.10 \pm 0.06$ & $0.08 \pm 0.05$ \\[1ex]
& Satellite & $0.10 \pm 0.07$ & $0.12 \pm 0.08$ & $0.09 \pm 0.06$ & $0.08 \pm 0.07$ \\[2ex]
Light-Weighted $[Z/H]$ & Central & $-0.14 \pm 0.05$  & $-0.10 \pm 0.06$  & $-0.10 \pm 0.06$ & $-0.11\pm 0.05$ \\[1ex]
& Satellite &  $-0.15 \pm 0.07$ & $-0.11 \pm 0.08$ & $-0.13 \pm 0.06$ &  $-0.11 \pm 0.07$ \\[2ex]
Mass-Weighted $[Z/H]$ & Central & $-0.03 \pm 0.05$ & $-0.07 \pm 0.06$ &  $-0.01 \pm 0.06$ & $-0.05 \pm 0.05$ \\[1ex]
& Satellite & $-0.09 \pm 0.07$ & $0.04 \pm 0.08$ &  $-0.05 \pm 0.06$ &  $0.04 \pm 0.07$ \\[2ex]
\hline
\end{tabular}
\label{table:cent_sat_table}
\end{table*}
Figure~\ref{fig:central_satellite} shows the light-weighted age and metallicity gradients for central (grey) and satellite (red) galaxies, as a function of local environment. Table~\ref{table:cent_sat_table} shows the numerical results of this analysis. First, Figure~\ref{fig:central_satellite} shows that stellar population gradients are independent of environmental density, as no correlation is evident between stellar population gradient and local density. This can be quantitively described by fitting a line through the stellar population gradient-environment plane in each panel plot. We find that luminosity and mass-weighted stellar population gradients generally do not correlate with local environment neither for central nor satellite galaxies. We further do not detect any evidence for a difference in gradients between satellite and central galaxies (see also Table~\ref{table:cent_sat_table}). We conclude that the galaxy environment, whether measured as local environmental density or through central/satellite classification, does not appear to have any significant effect on age and metallicity gradients in galaxies. This result agrees well with a recent IFU study of nearby massive galaxies as part of the MASSIVE survey, where it is found that even at large radius, internal properties matter more than environment in determining star formation history \citep{greene2015}. 

\subsection{Environmental Trends}
\begin{figure*}
\includegraphics[width=0.9\linewidth]{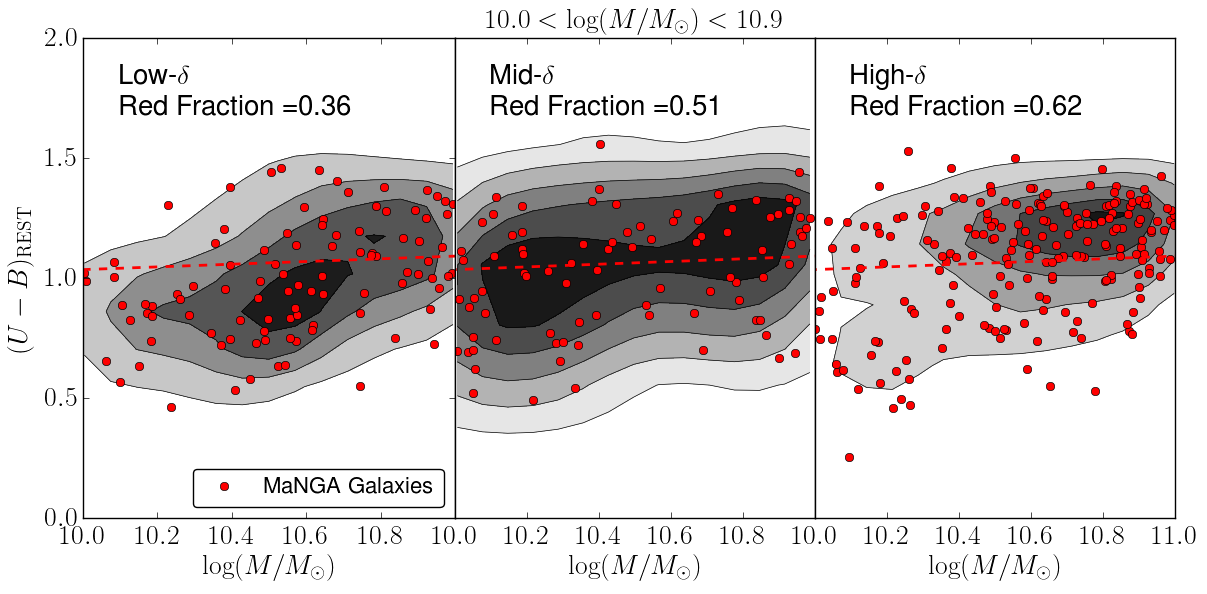}
\caption{Figure showing the colour-mass relation for three different environmental densities. The grey contours show the number density of points, the red circles show the MaNGA galaxies and the red dividing line shows the which distinguishes red and blue galaxies from \citet{peng2010}. The red fraction of galaxies is shown in each corresponding panel. All galaxies have a mass in the range $10 < \log(M/M_{\odot}) < 10.9$.}
\label{fig:red_fraction}
\end{figure*}
To ensure that our galaxy sample size was sufficient to identify different environmental impacts on gradients, we attempted to reproduce known environmental trends on galaxy properties. \citet{peng2010} studied the fraction of red galaxies as a function of environment and mass and found higher fractions of red galaxies exist in denser environments. We took a sample of galaxies our study ($10 < \log(M/M_{\odot}) < 10.9$), and calculated the red fraction for three different environment bins. The bins were defined in a similar fashion to what was done for the $N^{th}$ nearest neighbour, using percentiles of the environment distribution. Firstly, the $(u-g)_{\mathrm{REST}}$ colour for each galaxy was calculated using:
\begin{equation}
(u-g)_{\mathrm{REST}} = (u-g)-k_{ug}
\end{equation}
where $k_{ug}$ is the K-correction, which is small for the low redshift galaxy sample used in this work ($k_{ug} \approx 0.05$ magnitudes). Secondly, we used the transform equation of Lupton (2005), found on the SDSS website, to get $(U-B)_{\mathrm{REST}}$ colours.
\begin{equation}
(U-B)_{\mathrm{REST}} = 0.8116((u-g)_{\mathrm{REST}})-0.1313.
\end{equation}
Lastly, following the prescription of \citet{peng2010}, a dividing line was then employed so that we could define different galaxy populations. The dividing line has the form:
\begin{equation}
(U-B)_{\mathrm{REST}} = 1.10+0.075\log(m/10^{10} M_{\odot}) -0.182.
\end{equation}
Galaxies with $(U-B)_{\mathrm{REST}}$ greater than this were classed as red, and galaxies under this line were classed as blue (see Figure~\ref{fig:red_fraction}). We find that in the lowest density environments, the red fraction is $36\%$ and then increases up to $50\%$ in the next environmental density. In the highest density environment the fraction of red galaxies increases to $60\%$. This trend is similar to what was found in \citet{peng2010}. It is reassuring that environmental effects can be detected with the present density estimates and sample size. Hence any significant trends between stellar population gradient and environment are detectable with the present sample, and if there are any residual dependencies of stellar population gradients on environment, they must be a very subtle.

\section{Discussion} 
If the environment in which a galaxy resides has any significant impact over the timescales of gas dissipation and star formation, we might expect to see an environmental dependence on the inferred radial gradients presented in this work. Figure~\ref{fig:early_age_grad} and Table~\ref{table:complete_table} show the stellar population gradients obtained for four different environmental densities using the $N^{th}$ nearest neighbour method. For both early and late-type galaxies, the gradients do not vary much from one environmental density to another. This suggests that internal processes, such as supernova and Active Galactic Nuclei (AGN) feedback, matter most in determining the stellar population gradients in galaxies. Figure~\ref{fig:q_gradients}, which shows the stellar population gradients as a function of environment using the mass-dependant parameter $Q$, also corroborates with this view, as the gradients are relatively homogeneous across the $Q$ spectrum. 
\\
\\
In the parallel paper of \citet{zheng2016} the same lack of environmental dependence was found, agreeing with what is presented here. They find that disk galaxies have negative age and metallicity gradients, and elliptical galaxies have flat age gradients and negative metallicity gradients, qualitatively agreeing with our gradient values. These gradient values also remain consistent between the cluster, filament, sheet and void classification, showing no impact of environment. It is reassuring to see that a study using different methods, such as environment classification, full spectral fitting code and stellar population models, can produce similar conclusions to what is presented in this study.  
\\
\\
Another way of investigating environmental effects on stellar population gradients is to look at the difference between central and satellite galaxies, as these will be exposed to numerous different physical processes that can influence their evolution. Figure~\ref{fig:central_satellite} shows the gradients obtained for central and satellite galaxies as a function of $N^{th}$ nearest neighbour environmental density. We find that both central and satellite galaxies have relatively flat age gradients and negative metallicity gradients. This highlights the importance of internal properties, as opposed to location in the dark matter halo, on the inferred radial gradients. Table~\ref{table:cent_sat_table} shows the gradient values for the central and satellite galaxies as a function of four different environmental densities, and once again, no significant trend of the gradients of central and satellite galaxies with local environment is present. A study by \cite{brough2007} found similar results when investigating a sample of Brightest Group Galaxies (BGGs) and Brightest Cluster Galaxies (BCGs). 
\\
\\
Our results are also in reasonable agreement with a previous photometric study by \cite{tortora2012}, who analyse the differences in colour and stellar population gradients as a function of environment for central and satellite galaxies from SDSS imaging. They find that in most cases, age and metallicity gradients generally do not depend on environmental density. However, a mild residual dependence of metallicity gradient with environment is seen for central galaxies only, a pattern not detected here. This mild residual dependence has also been found in studies by \citet{blazquez2006}, using 82 galaxies in the coma cluster, and \citet{lababera2011} who used optical and near-infrared colours to study group and field galaxies. It will be interesting in future to see whether such a residual dependence can be recovered with larger MaNGA galaxy samples or alternative methodologies in future studies.

\section{Conclusions}
Mapping Nearby Galaxies at Apache Point Observatory (MaNGA) is a 6-year SDSS-IV survey that is obtaining spatially resolved spectroscopy for a sample of 10,000 nearby galaxies. In this paper, we study the internal gradients of stellar population properties, such as age and metallicity within $1.5\;R_{\rm e}$, for a representative sample of 721 galaxies taken from the first year of MaNGA observations (MPL4, equivalent to DR13) with masses ranging from $10^{9}\;M_{\odot}$ to $10^{11.5}\;M_{\odot}$. We split our galaxy sample into 505 early and 216 late-type galaxies based upon Galaxy Zoo classifications and analyse the impact of galaxy environment on the stellar population gradients. We calculate local environmental densities from the SDSS parent catalogue using $N^{th}$ nearest neighbour. In addition to this, we also look at a mass-dependent environmental measure, $Q$, which quantifies the tidal strength of nearest neighbours and split the MaNGA sample into central and satellite galaxies.\\
\\
We then apply the full spectral fitting code FIREFLY on these spectra to derive the stellar population parameters averaged age and metallicity. We use the stellar population models of \cite{maraston2011} (M11), which utilise the MILES stellar library \citep{miles2006} and assume a Kroupa stellar initial mass function (IMF, \cite{kroupa2011}). In our analysis, we find that early-type galaxies generally exhibit shallow light-weighted age gradients in agreement with the literature. However, the mass-weighted median age does show some radial dependence with positive gradients. Late-type galaxies, instead, have negative light-weighted age gradients in agreement with the literature. We generally detect negative metallicity gradients for both early and late-types at all masses, but these are significantly steeper in late-type compared to early-type galaxies. \\
\\
To understand the impact of galaxy environment on stellar population gradients, the galaxy sample was further split into four different local environmental densities. Distributions of age and metallicity gradients turn out to be indistinguishable across the different environments, and we also do not find any correlation between stellar population gradient and local density. K-S tests were conducted to confirm this result for both early and late-type galaxies. In addition to this, we repeated our analysis using the tidal strength parameter $Q$. This mass-dependent environment measure yielded similar results and the gradients appear to be indistinguishable across the different environments. We also split the sample into central and satellite galaxies and found that both the light and mass-weighted age and metallicity gradients are the same for both classes, and their values also do not vary across different environments. We therefore conclude that galaxy environment has no significant effect on age or metallicity gradients in galaxies at least within $1.5\;R_{\rm e}$, independently of mass or type. Hydrodynamical simulations of galaxy formation from the literature predict age gradients in early-type galaxies to be generally flat and independent of galaxy mass or environment, which agrees well with the findings of this paper. However, galaxy formation simulations seem to predict a dependence of metallicity gradients on environment, which is not confirmed by the results of the present study. A more comprehensive and direct comparison between MaNGA observations and predictions from galaxy formation simulations will be very valuable in future.

\section*{Acknowledgements}
The authors would like to thank Alfonso Aragon-Salamanca, Matthew Withers, Xan Morice-Atkinson for fruitful discussions and Francesco Belfiore for assisting with the Central/Satellite galaxy catalogue. DG is supported by an STFC PhD studentship. MAB acknowledges NSF AST-1517006. AW acknowledges support from a Leverhulme Early Career Fellowship. Numerical computations were done on the Sciama High Performance Compute (HPC) cluster which is supported by the Institute of Cosmology of Gravitation, SEPNet and the University of Portsmouth. Funding for the Sloan Digital Sky Survey IV has been provided by the Alfred P. Sloan Foundation, the U.S. Department of Energy Office of Science, and the Participating Institutions. SDSS- IV acknowledges support and resources from the Center for High-Performance Computing at the University of Utah. The SDSS web site is www.sdss.org. SDSS-IV is managed by the Astrophysical Research Consortium for the Participating Institutions of the SDSS Collaboration including the Brazilian Participation Group, the Carnegie Institution for Science, Carnegie Mellon University, the Chilean Participation Group, the French Participation Group, Harvard-Smithsonian Center for Astrophysics, Instituto de Astrofísica de Canarias, The Johns Hopkins University, Kavli Institute for the Physics and Mathematics of the Universe (IPMU) / University of Tokyo, Lawrence Berkeley National Laboratory, Leibniz Institut für Astrophysik Potsdam (AIP), Max-Planck-Institut für Astronomie (MPIA Heidelberg), Max-Planck-Institut für Astrophysik (MPA Garching), Max-Planck-Institut für Extraterrestrische Physik (MPE), National Astronomical Observatory of China, New Mexico State University, New York University, University of Notre Dame, Observatório Nacional / MCTI, The Ohio State University, Pennsylvania State University, Shanghai Astronomical Observatory, United Kingdom Participation Group, Universidad Nacional Autónoma de México, University of Arizona, University of Colorado Boulder, University of Oxford, University of Portsmouth, University of Utah, University of Virginia, University of Washington, University of Wisconsin, Vanderbilt University, and Yale University. \\
\\
{\it All data taken as part of SDSS-IV is scheduled to be released to the community in fully reduced form at regular intervals through dedicated data releases. The first MaNGA data release was part of the SDSS data release 13 (release date 31 July 2016).}
\\
\\
 $^{1}$Institute of Cosmology and Gravitation, University of Portsmouth, Burnaby Road, Portsmouth, UK, PO1 3FX.\\
$^{2}$Instituto de Física, Universidade Federal do Rio Grande do Sul, Campus do Vale, Porto Alegre, Brasil.\\
$^{3}$Laboratório Interinstitucional de e- Astronomia, Rua General José Cristino, 77 Vasco da Gama, Rio de Janeiro, Brasil.\\
$^{4}$National Astronomical Observatories, Chinese Academy of Sciences, A20 Datun Road, Chaoyang District, Beijing 100012, China.\\
$^{5}$Unidad de Astronomía, Fac. Cs. Básicas, U. de Antofagasta, Avda. U. de Antofagasta 02800, Antofagasta, Chile.\\
$^{6}$University of Wisconsin-Madison, Department of Astronomy, 475 N. Charter Street, Madison, WI 53706-1582, USA. \\
$^{7}$Kavli InstitGute for the Physics and Mathematics of the Universe (WPI), The University of Tokyo Institutes for Advanced Study, Kashiwa, Chiba 277-8583, Japan.\\
$^{8}$McDonald Observatory, Department of Astronomy, University of Texas at Austin, 1 University Station, Austin, TX 78712- 0259, USA.\\
$^{9}$Space Telescope Science Institute, 3700 San Martin Drive, Baltimore, MD 21218, USA.\\
$^{10}$Department of Physics and Astronomy, University of Kentucky, 505 Rose St., Lexington, KY 40506-0057, USA.\\
$^{11}$Department of Physical Sciences, The Open University, Milton Keynes, UK.\\
$^{12}$School of Physics and Astronomy, University of St. Andrews, North Haugh, St. Andrews, KY16 9SS, UK.\\
$^{13}$Apache Point Observatory, P.O. Box 59, Sunspot, NM 88349, USA.\\
$^{14}$Sternberg Astronomical Institute, Moscow State University, Moscow, Russia. \\
$^{15}$Department of Physics and Astronomy, University of Utah, 115 S. 1400 E., Salt Lake City, UT 84112, USA.\\
$^{16}$Instituto de Astrof{\'i}sica, Pontificia Universidad Católica de Chile, Av. Vicuna Mackenna 4860, 782-0436 Macul, Santiago, Chile.\\
$^{17}$University of Cambridge, Cavendish Astrophysics, Cambridge, CB3 0HE, UK.\\
$^{18}$University of Cambridge, Kavli Institute for Cosmology, Cambridge, CB3 0HE, UK.\\
$^{19}$School of Physics and Astronomy, University of Nottingham, University Park, Nottingham NG7 2RD, UK.\\
$^{20}$Unidad de Astronomía, Universidad de Antofagasta, Avenida Angamos 601, Antofagasta 1270300, Chile.\\
$^{21}$Departamento de F{\'i}sica, Facultad de Ciencias, Universidad de La Serena, Cisternas 1200, La Serena, Chile. 

\bibliography{test_paper_bib}

\begin{thebibliography}{}
\makeatletter
\relax
\def\mn@urlcharsother{\let\do\@makeother \do\$\do\&\do\#\do\^\do\_\do\%\do\~}
\def\mn@doi{\begingroup\mn@urlcharsother \@ifnextchar [ {\mn@doi@}
  {\mn@doi@[]}}
\def\mn@doi@[#1]#2{\def\@tempa{#1}\ifx\@tempa\@empty \href
  {http://dx.doi.org/#2} {doi:#2}\else \href {http://dx.doi.org/#2} {#1}\fi
  \endgroup}
\def\mn@eprint#1#2{\mn@eprint@#1:#2::\@nil}
\def\mn@eprint@arXiv#1{\href {http://arxiv.org/abs/#1} {{\tt arXiv:#1}}}
\def\mn@eprint@dblp#1{\href {http://dblp.uni-trier.de/rec/bibtex/#1.xml}
  {dblp:#1}}
\def\mn@eprint@#1:#2:#3:#4\@nil{\def\@tempa {#1}\def\@tempb {#2}\def\@tempc
  {#3}\ifx \@tempc \@empty \let \@tempc \@tempb \let \@tempb \@tempa \fi \ifx
  \@tempb \@empty \def\@tempb {arXiv}\fi \@ifundefined
  {mn@eprint@\@tempb}{\@tempb:\@tempc}{\expandafter \expandafter \csname
  mn@eprint@\@tempb\endcsname \expandafter{\@tempc}}}

\bibitem[\protect\citeauthoryear{{Alam} et~al.,}{{Alam} et~al.}{2015}]{dr12}
{Alam} S.,  et~al., 2015, \mn@doi [\apjs] {10.1088/0067-0049/219/1/12}, \href
  {http://adsabs.harvard.edu/abs/2015ApJS..219...12A} {219, 12}

\bibitem[\protect\citeauthoryear{{Allen} et~al.,}{{Allen}
  et~al.}{2015}]{allen2015}
{Allen} J.~T.,  et~al., 2015, \mn@doi [\mnras] {10.1093/mnras/stu2057}, \href
  {http://adsabs.harvard.edu/abs/2015MNRAS.446.1567A} {446, 1567}

\bibitem[\protect\citeauthoryear{{Argudo-Fern{\'a}ndez}
  et~al.,}{{Argudo-Fern{\'a}ndez} et~al.}{2013}]{maria1}
{Argudo-Fern{\'a}ndez} M.,  et~al., 2013, \mn@doi [\aap]
  {10.1051/0004-6361/201321326}, \href
  {http://adsabs.harvard.edu/abs/2013A%26A...560A...9A} {560, A9}

\bibitem[\protect\citeauthoryear{{Argudo-Fern{\'a}ndez}
  et~al.,}{{Argudo-Fern{\'a}ndez} et~al.}{2014}]{maria2}
{Argudo-Fern{\'a}ndez} M.,  et~al., 2014, \mn@doi [\aap]
  {10.1051/0004-6361/201322498}, \href
  {http://adsabs.harvard.edu/abs/2014A%26A...564A..94A} {564, A94}

\bibitem[\protect\citeauthoryear{{Argudo-Fern{\'a}ndez}
  et~al.,}{{Argudo-Fern{\'a}ndez} et~al.}{2015}]{maria3}
{Argudo-Fern{\'a}ndez} M.,  et~al., 2015, \mn@doi [\aap]
  {10.1051/0004-6361/201526016}, \href
  {http://adsabs.harvard.edu/abs/2015A%26A...578A.110A} {578, A110}

\bibitem[\protect\citeauthoryear{{Bacon} et~al.,}{{Bacon}
  et~al.}{1995}]{bacon1995}
{Bacon} R.,  et~al., 1995, \aaps, \href
  {http://adsabs.harvard.edu/abs/1995A%26AS..113..347B} {113, 347}

\bibitem[\protect\citeauthoryear{{Baldry}, {Balogh}, {Bower}, {Glazebrook},
  {Nichol}, {Bamford}  \& {Budavari}}{{Baldry} et~al.}{2006}]{baldry2007}
{Baldry} I.~K.,  {Balogh} M.~L.,  {Bower} R.~G.,  {Glazebrook} K.,  {Nichol}
  R.~C.,  {Bamford} S.~P.,   {Budavari} T.,  2006, \mn@doi [\mnras]
  {10.1111/j.1365-2966.2006.11081.x}, \href
  {http://adsabs.harvard.edu/abs/2006MNRAS.373..469B} {373, 469}

\bibitem[\protect\citeauthoryear{{Bell}, {McIntosh}, {Katz}  \&
  {Weinberg}}{{Bell} et~al.}{2003}]{bell2003}
{Bell} E.~F.,  {McIntosh} D.~H.,  {Katz} N.,   {Weinberg} M.~D.,  2003, \mn@doi
  [\apjs] {10.1086/378847}, \href
  {http://adsabs.harvard.edu/abs/2003ApJS..149..289B} {149, 289}

\bibitem[\protect\citeauthoryear{{Bell}, {Phleps}, {Somerville}, {Wolf},
  {Borch}  \& {Meisenheimer}}{{Bell} et~al.}{2006}]{bell2006}
{Bell} E.~F.,  {Phleps} S.,  {Somerville} R.~S.,  {Wolf} C.,  {Borch} A.,
  {Meisenheimer} K.,  2006, \mn@doi [\apj] {10.1086/508408}, \href
  {http://adsabs.harvard.edu/abs/2006ApJ...652..270B} {652, 270}

\bibitem[\protect\citeauthoryear{{Bernstein} \& {Bhavsar}}{{Bernstein} \&
  {Bhavsar}}{2001}]{bern2001}
{Bernstein} J.~P.,  {Bhavsar} S.~P.,  2001, \mn@doi [\mnras]
  {10.1046/j.1365-8711.2001.04124.x}, \href
  {http://adsabs.harvard.edu/abs/2001MNRAS.322..625B} {322, 625}

\bibitem[\protect\citeauthoryear{{Bershady}, {Verheijen}, {Swaters},
  {Andersen}, {Westfall}  \& {Martinsson}}{{Bershady}
  et~al.}{2010}]{bershady2010}
{Bershady} M.~A.,  {Verheijen} M.~A.~W.,  {Swaters} R.~A.,  {Andersen} D.~R.,
  {Westfall} K.~B.,   {Martinsson} T.,  2010, \mn@doi [\apj]
  {10.1088/0004-637X/716/1/198}, \href
  {http://adsabs.harvard.edu/abs/2010ApJ...716..198B} {716, 198}

\bibitem[\protect\citeauthoryear{{Bershady}, {Martinsson}, {Verheijen},
  {Westfall}, {Andersen}  \& {Swaters}}{{Bershady} et~al.}{2011}]{bershady2011}
{Bershady} M.~A.,  {Martinsson} T.~P.~K.,  {Verheijen} M.~A.~W.,  {Westfall}
  K.~B.,  {Andersen} D.~R.,   {Swaters} R.~A.,  2011, \mn@doi [\apjl]
  {10.1088/2041-8205/739/2/L47}, \href
  {http://adsabs.harvard.edu/abs/2011ApJ...739L..47B} {739, L47}

\bibitem[\protect\citeauthoryear{{Blanton} \& {Moustakas}}{{Blanton} \&
  {Moustakas}}{2009}]{blanton2009}
{Blanton} M.~R.,  {Moustakas} J.,  2009, \mn@doi [\araa]
  {10.1146/annurev-astro-082708-101734}, \href
  {http://adsabs.harvard.edu/abs/2009ARA%26A..47..159B} {47, 159}

\bibitem[\protect\citeauthoryear{{Blanton} et~al.,}{{Blanton}
  et~al.}{2005a}]{blantoncat}
{Blanton} M.~R.,  et~al., 2005a, \mn@doi [\aj] {10.1086/429803}, \href
  {http://adsabs.harvard.edu/abs/2005AJ....129.2562B} {129, 2562}

\bibitem[\protect\citeauthoryear{{Blanton}, {Eisenstein}, {Hogg}, {Schlegel}
  \& {Brinkmann}}{{Blanton} et~al.}{2005b}]{blanton2005}
{Blanton} M.~R.,  {Eisenstein} D.,  {Hogg} D.~W.,  {Schlegel} D.~J.,
  {Brinkmann} J.,  2005b, \mn@doi [\apj] {10.1086/422897}, \href
  {http://adsabs.harvard.edu/abs/2005ApJ...629..143B} {629, 143}

\bibitem[\protect\citeauthoryear{{Brough}, {Proctor}, {Forbes}, {Couch},
  {Collins}, {Burke}  \& {Mann}}{{Brough} et~al.}{2007}]{brough2007}
{Brough} S.,  {Proctor} R.,  {Forbes} D.~A.,  {Couch} W.~J.,  {Collins} C.~A.,
  {Burke} D.~J.,   {Mann} R.~G.,  2007, \mn@doi [\mnras]
  {10.1111/j.1365-2966.2007.11900.x}, \href
  {http://adsabs.harvard.edu/abs/2007MNRAS.378.1507B} {378, 1507}

\bibitem[\protect\citeauthoryear{{Bruzual} \& {Charlot}}{{Bruzual} \&
  {Charlot}}{2003}]{bruzual2003}
{Bruzual} G.,  {Charlot} S.,  2003, \mn@doi [\mnras]
  {10.1046/j.1365-8711.2003.06897.x}, \href
  {http://adsabs.harvard.edu/abs/2003MNRAS.344.1000B} {344, 1000}

\bibitem[\protect\citeauthoryear{{Bundy} et~al.,}{{Bundy}
  et~al.}{2015}]{bundy2015}
{Bundy} K.,  et~al., 2015, \mn@doi [\apj] {10.1088/0004-637X/798/1/7}, \href
  {http://adsabs.harvard.edu/abs/2015ApJ...798....7B} {798, 7}

\bibitem[\protect\citeauthoryear{{Cappellari} \& {Emsellem}}{{Cappellari} \&
  {Emsellem}}{2004}]{emsellem}
{Cappellari} M.,  {Emsellem} E.,  2004, \mn@doi [\pasp] {10.1086/381875}, \href
  {http://adsabs.harvard.edu/abs/2004PASP..116..138C} {116, 138}

\bibitem[\protect\citeauthoryear{{Cappellari} et~al.,}{{Cappellari}
  et~al.}{2011}]{cappellari2011}
{Cappellari} M.,  et~al., 2011, \mn@doi [\mnras]
  {10.1111/j.1365-2966.2010.18174.x}, \href
  {http://adsabs.harvard.edu/abs/2011MNRAS.413..813C} {413, 813}

\bibitem[\protect\citeauthoryear{{Chabrier}}{{Chabrier}}{2003}]{chabrier2003}
{Chabrier} G.,  2003, \mn@doi [\pasp] {10.1086/376392}, \href
  {http://adsabs.harvard.edu/abs/2003PASP..115..763C} {115, 763}

\bibitem[\protect\citeauthoryear{{Cid Fernandes}, {Mateus}, {Sodr{\'e}},
  {Stasi{\'n}ska}  \& {Gomes}}{{Cid Fernandes} et~al.}{2005}]{starlight2005}
{Cid Fernandes} R.,  {Mateus} A.,  {Sodr{\'e}} L.,  {Stasi{\'n}ska} G.,
  {Gomes} J.~M.,  2005, \mn@doi [\mnras] {10.1111/j.1365-2966.2005.08752.x},
  \href {http://adsabs.harvard.edu/abs/2005MNRAS.358..363C} {358, 363}

\bibitem[\protect\citeauthoryear{{Colless} et~al.,}{{Colless}
  et~al.}{2001}]{colless2003}
{Colless} M.,  et~al., 2001, \mn@doi [\mnras]
  {10.1046/j.1365-8711.2001.04902.x}, \href
  {http://adsabs.harvard.edu/abs/2001MNRAS.328.1039C} {328, 1039}

\bibitem[\protect\citeauthoryear{{Cooper}, {Newman}, {Madgwick}, {Gerke}, {Yan}
   \& {Davis}}{{Cooper} et~al.}{2005}]{cooper2005}
{Cooper} M.~C.,  {Newman} J.~A.,  {Madgwick} D.~S.,  {Gerke} B.~F.,  {Yan} R.,
   {Davis} M.,  2005, \mn@doi [\apj] {10.1086/432868}, \href
  {http://adsabs.harvard.edu/abs/2005ApJ...634..833C} {634, 833}

\bibitem[\protect\citeauthoryear{{Davies}, {M{\"u}ller S{\'a}nchez}, {Genzel},
  {Tacconi}, {Hicks}, {Friedrich}  \& {Sternberg}}{{Davies}
  et~al.}{2007}]{ricdavies2007}
{Davies} R.~I.,  {M{\"u}ller S{\'a}nchez} F.,  {Genzel} R.,  {Tacconi} L.~J.,
  {Hicks} E.~K.~S.,  {Friedrich} S.,   {Sternberg} A.,  2007, \mn@doi [\apj]
  {10.1086/523032}, \href {http://adsabs.harvard.edu/abs/2007ApJ...671.1388D}
  {671, 1388}

\bibitem[\protect\citeauthoryear{{Davis}, {Efstathiou}, {Frenk}  \&
  {White}}{{Davis} et~al.}{1985}]{white1985}
{Davis} M.,  {Efstathiou} G.,  {Frenk} C.~S.,   {White} S.~D.~M.,  1985,
  \mn@doi [\apj] {10.1086/163168}, \href
  {http://adsabs.harvard.edu/abs/1985ApJ...292..371D} {292, 371}

\bibitem[\protect\citeauthoryear{{Dressler}}{{Dressler}}{1980}]{dressler1980}
{Dressler} A.,  1980, \mn@doi [\apj] {10.1086/157753}, \href
  {http://adsabs.harvard.edu/abs/1980ApJ...236..351D} {236, 351}

\bibitem[\protect\citeauthoryear{{Eardley} et~al.,}{{Eardley}
  et~al.}{2015}]{eardley2015}
{Eardley} E.,  et~al., 2015, \mn@doi [\mnras] {10.1093/mnras/stv237}, \href
  {http://adsabs.harvard.edu/abs/2015MNRAS.448.3665E} {448, 3665}

\bibitem[\protect\citeauthoryear{{Etherington} \& {Thomas}}{{Etherington} \&
  {Thomas}}{2015}]{etherington2015}
{Etherington} J.,  {Thomas} D.,  2015, \mn@doi [\mnras] {10.1093/mnras/stv999},
  \href {http://adsabs.harvard.edu/abs/2015MNRAS.451..660E} {451, 660}

\bibitem[\protect\citeauthoryear{{Farouki} \& {Shapiro}}{{Farouki} \&
  {Shapiro}}{1981}]{farouki1981}
{Farouki} R.,  {Shapiro} S.~L.,  1981, \mn@doi [\apj] {10.1086/158563}, \href
  {http://adsabs.harvard.edu/abs/1981ApJ...243...32F} {243, 32}

\bibitem[\protect\citeauthoryear{{Fitzpatrick}}{{Fitzpatrick}}{1999}]{fitzpatrick1999}
{Fitzpatrick} E.~L.,  1999, \mn@doi [\pasp] {10.1086/316293}, \href
  {http://adsabs.harvard.edu/abs/1999PASP..111...63F} {111, 63}

\bibitem[\protect\citeauthoryear{{Goddard} et~al.}{{Goddard}
  et~al.}{2016}]{goddard2016a}
{Goddard} D.,  et~al., 2016, \mnras, submitted

\bibitem[\protect\citeauthoryear{{Gonz{\'a}lez Delgado} et~al.,}{{Gonz{\'a}lez
  Delgado} et~al.}{2015}]{gonz2015}
{Gonz{\'a}lez Delgado} R.~M.,  et~al., 2015, \mn@doi [\aap]
  {10.1051/0004-6361/201525938}, \href
  {http://adsabs.harvard.edu/abs/2015A%26A...581A.103G} {581, A103}

\bibitem[\protect\citeauthoryear{{Greene}, {Janish}, {Ma}, {McConnell},
  {Blakeslee}, {Thomas}  \& {Murphy}}{{Greene} et~al.}{2015}]{greene2015}
{Greene} J.~E.,  {Janish} R.,  {Ma} C.-P.,  {McConnell} N.~J.,  {Blakeslee}
  J.~P.,  {Thomas} J.,   {Murphy} J.~D.,  2015, \mn@doi [\apj]
  {10.1088/0004-637X/807/1/11}, \href
  {http://adsabs.harvard.edu/abs/2015ApJ...807...11G} {807, 11}

\bibitem[\protect\citeauthoryear{{Guth}}{{Guth}}{1981}]{guth1981}
{Guth} A.~H.,  1981, \mn@doi [\prd] {10.1103/PhysRevD.23.347}, \href
  {http://adsabs.harvard.edu/abs/1981PhRvD..23..347G} {23, 347}

\bibitem[\protect\citeauthoryear{{Hahn}, {Carollo}, {Porciani}  \&
  {Dekel}}{{Hahn} et~al.}{2007}]{hahn2007}
{Hahn} O.,  {Carollo} C.~M.,  {Porciani} C.,   {Dekel} A.,  2007, \mn@doi
  [\mnras] {10.1111/j.1365-2966.2007.12249.x}, \href
  {http://adsabs.harvard.edu/abs/2007MNRAS.381...41H} {381, 41}

\bibitem[\protect\citeauthoryear{{Hirschmann}, {Naab}, {Ostriker}, {Forbes},
  {Duc}, {Dav{\'e}}, {Oser}  \& {Karabal}}{{Hirschmann}
  et~al.}{2015}]{hirschmann2015}
{Hirschmann} M.,  {Naab} T.,  {Ostriker} J.~P.,  {Forbes} D.~A.,  {Duc} P.-A.,
  {Dav{\'e}} R.,  {Oser} L.,   {Karabal} E.,  2015, \mn@doi [\mnras]
  {10.1093/mnras/stv274}, \href
  {http://adsabs.harvard.edu/abs/2015MNRAS.449..528H} {449, 528}

\bibitem[\protect\citeauthoryear{{Hogg} et~al.,}{{Hogg}
  et~al.}{2004}]{hogg2004}
{Hogg} D.~W.,  et~al., 2004, \mn@doi [\apjl] {10.1086/381749}, \href
  {http://adsabs.harvard.edu/abs/2004ApJ...601L..29H} {601, L29}

\bibitem[\protect\citeauthoryear{{Kamann} et~al.,}{{Kamann}
  et~al.}{2016}]{kaman2016}
{Kamann} S.,  et~al., 2016, The Messenger, \href
  {http://esoads.eso.org/abs/2016Msngr.164...18K} {164, 18}

\bibitem[\protect\citeauthoryear{{Kauffmann}, {White}, {Heckman}, {M{\'e}nard},
  {Brinchmann}, {Charlot}, {Tremonti}  \& {Brinkmann}}{{Kauffmann}
  et~al.}{2004}]{kauffmann2004}
{Kauffmann} G.,  {White} S.~D.~M.,  {Heckman} T.~M.,  {M{\'e}nard} B.,
  {Brinchmann} J.,  {Charlot} S.,  {Tremonti} C.,   {Brinkmann} J.,  2004,
  \mn@doi [\mnras] {10.1111/j.1365-2966.2004.08117.x}, \href
  {http://adsabs.harvard.edu/abs/2004MNRAS.353..713K} {353, 713}

\bibitem[\protect\citeauthoryear{{Kroupa}}{{Kroupa}}{2001}]{kroupa2011}
{Kroupa} P.,  2001, \mn@doi [\mnras] {10.1046/j.1365-8711.2001.04022.x}, \href
  {http://adsabs.harvard.edu/abs/2001MNRAS.322..231K} {322, 231}

\bibitem[\protect\citeauthoryear{{Kuntschner} et~al.,}{{Kuntschner}
  et~al.}{2010}]{kuntschner2010}
{Kuntschner} H.,  et~al., 2010, \mn@doi [\mnras]
  {10.1111/j.1365-2966.2010.17161.x}, \href
  {http://adsabs.harvard.edu/abs/2010MNRAS.408...97K} {408, 97}

\bibitem[\protect\citeauthoryear{{La Barbera}, {Ferreras}, {de Carvalho},
  {Lopes}, {Pasquali}, {de la Rosa}  \& {De Lucia}}{{La Barbera}
  et~al.}{2011a}]{barbera2011}
{La Barbera} F.,  {Ferreras} I.,  {de Carvalho} R.~R.,  {Lopes} P.~A.~A.,
  {Pasquali} A.,  {de la Rosa} I.~G.,   {De Lucia} G.,  2011a, \mn@doi [\apjl]
  {10.1088/2041-8205/740/2/L41}, \href
  {http://adsabs.harvard.edu/abs/2011ApJ...740L..41L} {740, L41}

\bibitem[\protect\citeauthoryear{{La Barbera}, {Ferreras}, {de Carvalho},
  {Lopes}, {Pasquali}, {de la Rosa}  \& {De Lucia}}{{La Barbera}
  et~al.}{2011b}]{lababera2011}
{La Barbera} F.,  {Ferreras} I.,  {de Carvalho} R.~R.,  {Lopes} P.~A.~A.,
  {Pasquali} A.,  {de la Rosa} I.~G.,   {De Lucia} G.,  2011b, \mn@doi [\apjl]
  {10.1088/2041-8205/740/2/L41}, \href
  {http://adsabs.harvard.edu/abs/2011ApJ...740L..41L} {740, L41}

\bibitem[\protect\citeauthoryear{{Larson}, {Tinsley}  \& {Caldwell}}{{Larson}
  et~al.}{1980}]{larson1980}
{Larson} R.~B.,  {Tinsley} B.~M.,   {Caldwell} C.~N.,  1980, \mn@doi [\apj]
  {10.1086/157917}, \href {http://adsabs.harvard.edu/abs/1980ApJ...237..692L}
  {237, 692}

\bibitem[\protect\citeauthoryear{{Law} et~al.,}{{Law} et~al.}{2015}]{law2015}
{Law} D.~R.,  et~al., 2015, \mn@doi [\aj] {10.1088/0004-6256/150/1/19}, \href
  {http://adsabs.harvard.edu/abs/2015AJ....150...19L} {150, 19}

\bibitem[\protect\citeauthoryear{{Law} et~al.}{{Law} et~al.}{2016}]{law2016}
{Law} D.,  et~al., 2016, \apj, in press

\bibitem[\protect\citeauthoryear{{Liddle}}{{Liddle}}{2007}]{liddle2007}
{Liddle} A.~R.,  2007, \mn@doi [\mnras] {10.1111/j.1745-3933.2007.00306.x},
  \href {http://adsabs.harvard.edu/abs/2007MNRAS.377L..74L} {377, L74}

\bibitem[\protect\citeauthoryear{{Lintott} et~al.,}{{Lintott}
  et~al.}{2011}]{lintott2011}
{Lintott} C.,  et~al., 2011, \mn@doi [\mnras]
  {10.1111/j.1365-2966.2010.17432.x}, \href
  {http://adsabs.harvard.edu/abs/2011MNRAS.410..166L} {410, 166}

\bibitem[\protect\citeauthoryear{{Maraston} \& {Str{\"o}mb{\"a}ck}}{{Maraston}
  \& {Str{\"o}mb{\"a}ck}}{2011}]{maraston2011}
{Maraston} C.,  {Str{\"o}mb{\"a}ck} G.,  2011, \mn@doi [\mnras]
  {10.1111/j.1365-2966.2011.19738.x}, \href
  {http://adsabs.harvard.edu/abs/2011MNRAS.418.2785M} {418, 2785}

\bibitem[\protect\citeauthoryear{{McDermid} et~al.,}{{McDermid}
  et~al.}{2006}]{mcdermid2006}
{McDermid} R.~M.,  et~al., 2006, \mn@doi [\nar] {10.1016/j.newar.2005.10.025},
  \href {http://adsabs.harvard.edu/abs/2006NewAR..49..521M} {49, 521}

\bibitem[\protect\citeauthoryear{{Mehlert}, {Thomas}, {Saglia}, {Bender}  \&
  {Wegner}}{{Mehlert} et~al.}{2003}]{mehlert2003}
{Mehlert} D.,  {Thomas} D.,  {Saglia} R.~P.,  {Bender} R.,   {Wegner} G.,
  2003, \mn@doi [\aap] {10.1051/0004-6361:20030886}, \href
  {http://adsabs.harvard.edu/abs/2003A%26A...407..423M} {407, 423}

\bibitem[\protect\citeauthoryear{{Mercurio}, {Haines}, {Gargiulo}, {La
  Barbera}, {Merluzzi}  \& {Busarello}}{{Mercurio} et~al.}{2010}]{mercurio2010}
{Mercurio} A.,  {Haines} C.~P.,  {Gargiulo} A.,  {La Barbera} F.,  {Merluzzi}
  P.,   {Busarello} G.,  2010, preprint, \href
  {http://adsabs.harvard.edu/abs/2010arXiv1006.5001M} {} (\mn@eprint {arXiv}
  {1006.5001})

\bibitem[\protect\citeauthoryear{{Muldrew} et~al.,}{{Muldrew}
  et~al.}{2012}]{muldrew2012}
{Muldrew} S.~I.,  et~al., 2012, \mn@doi [\mnras]
  {10.1111/j.1365-2966.2011.19922.x}, \href
  {http://adsabs.harvard.edu/abs/2012MNRAS.419.2670M} {419, 2670}

\bibitem[\protect\citeauthoryear{{Oemler}}{{Oemler}}{1974}]{oemler1974}
{Oemler} Jr. A.,  1974, \mn@doi [\apj] {10.1086/153216}, \href
  {http://adsabs.harvard.edu/abs/1974ApJ...194....1O} {194, 1}

\bibitem[\protect\citeauthoryear{{Peng} et~al.,}{{Peng}
  et~al.}{2010}]{peng2010}
{Peng} Y.-j.,  et~al., 2010, \mn@doi [\apj] {10.1088/0004-637X/721/1/193},
  \href {http://adsabs.harvard.edu/abs/2010ApJ...721..193P} {721, 193}

\bibitem[\protect\citeauthoryear{{P{\'e}rez} et~al.,}{{P{\'e}rez}
  et~al.}{2013}]{perez2013}
{P{\'e}rez} E.,  et~al., 2013, \mn@doi [\apjl] {10.1088/2041-8205/764/1/L1},
  \href {http://adsabs.harvard.edu/abs/2013ApJ...764L...1P} {764, L1}

\bibitem[\protect\citeauthoryear{{Planck Collaboration} et~al.,}{{Planck
  Collaboration} et~al.}{2015}]{planck2015}
{Planck Collaboration} et~al., 2015, preprint, \href
  {http://adsabs.harvard.edu/abs/2015arXiv150201589P} {} (\mn@eprint {arXiv}
  {1502.01589})

\bibitem[\protect\citeauthoryear{{Press}, {Teukolsky}, {Vetterling}  \&
  {Flannery}}{{Press} et~al.}{2007}]{recipes}
{Press} W.~H.,  {Teukolsky} S.~A.,  {Vetterling} W.~T.,   {Flannery} B.~P.,
  2007, {Numerical Recipes: The Art of Scientific Computing. Third Edition}.
Cambridge University Press, New York, NY USA

\bibitem[\protect\citeauthoryear{{Rawle}, {Smith}  \& {Lucey}}{{Rawle}
  et~al.}{2010}]{rawle2010}
{Rawle} T.~D.,  {Smith} R.~J.,   {Lucey} J.~R.,  2010, \mn@doi [\mnras]
  {10.1111/j.1365-2966.2009.15722.x}, \href
  {http://adsabs.harvard.edu/abs/2010MNRAS.401..852R} {401, 852}

\bibitem[\protect\citeauthoryear{{Read}, {Wilkinson}, {Evans}, {Gilmore}  \&
  {Kleyna}}{{Read} et~al.}{2006}]{read2006}
{Read} J.~I.,  {Wilkinson} M.~I.,  {Evans} N.~W.,  {Gilmore} G.,   {Kleyna}
  J.~T.,  2006, \mn@doi [\mnras] {10.1111/j.1365-2966.2005.09861.x}, \href
  {http://adsabs.harvard.edu/abs/2006MNRAS.366..429R} {366, 429}

\bibitem[\protect\citeauthoryear{{Riffel}, {Storchi-Bergmann}, {Riffel}  \&
  {Pastoriza}}{{Riffel} et~al.}{2010}]{brasil1}
{Riffel} R.~A.,  {Storchi-Bergmann} T.,  {Riffel} R.,   {Pastoriza} M.~G.,
  2010, \mn@doi [\apj] {10.1088/0004-637X/713/1/469}, \href
  {http://adsabs.harvard.edu/abs/2010ApJ...713..469R} {713, 469}

\bibitem[\protect\citeauthoryear{{Riffel}, {Riffel}, {Ferrari}  \&
  {Storchi-Bergmann}}{{Riffel} et~al.}{2011}]{brasil2}
{Riffel} R.,  {Riffel} R.~A.,  {Ferrari} F.,   {Storchi-Bergmann} T.,  2011,
  \mn@doi [\mnras] {10.1111/j.1365-2966.2011.19061.x}, \href
  {http://adsabs.harvard.edu/abs/2011MNRAS.416..493R} {416, 493}

\bibitem[\protect\citeauthoryear{{S{\'a}nchez-Bl{\'a}zquez}
  et~al.,}{{S{\'a}nchez-Bl{\'a}zquez} et~al.}{2006a}]{miles2006}
{S{\'a}nchez-Bl{\'a}zquez} P.,  et~al., 2006a, \mn@doi [\mnras]
  {10.1111/j.1365-2966.2006.10699.x}, \href
  {http://adsabs.harvard.edu/abs/2006MNRAS.371..703S} {371, 703}

\bibitem[\protect\citeauthoryear{{S{\'a}nchez-Bl{\'a}zquez}, {Gorgas}  \&
  {Cardiel}}{{S{\'a}nchez-Bl{\'a}zquez} et~al.}{2006b}]{blazquez2006}
{S{\'a}nchez-Bl{\'a}zquez} P.,  {Gorgas} J.,   {Cardiel} N.,  2006b, \mn@doi
  [\aap] {10.1051/0004-6361:20064846}, \href
  {http://adsabs.harvard.edu/abs/2006A%26A...457..823S} {457, 823}

\bibitem[\protect\citeauthoryear{{S{\'a}nchez-Bl{\'a}zquez}
  et~al.,}{{S{\'a}nchez-Bl{\'a}zquez} et~al.}{2014}]{sanchez2014}
{S{\'a}nchez-Bl{\'a}zquez} P.,  et~al., 2014, \mn@doi [\aap]
  {10.1051/0004-6361/201423635}, \href
  {http://adsabs.harvard.edu/abs/2014A%26A...570A...6S} {570, A6}

\bibitem[\protect\citeauthoryear{{S{\'a}nchez} et~al.,}{{S{\'a}nchez}
  et~al.}{2012}]{sanchez2012}
{S{\'a}nchez} S.~F.,  et~al., 2012, \mn@doi [\aap]
  {10.1051/0004-6361/201117353}, \href
  {http://adsabs.harvard.edu/abs/2012A%26A...538A...8S} {538, A8}

\bibitem[\protect\citeauthoryear{{Schawinski} et~al.,}{{Schawinski}
  et~al.}{2007}]{schawinski2007}
{Schawinski} K.,  et~al., 2007, \mn@doi [\apjs] {10.1086/516631}, \href
  {http://adsabs.harvard.edu/abs/2007ApJS..173..512S} {173, 512}

\bibitem[\protect\citeauthoryear{{Schlegel}, {Finkbeiner}  \&
  {Davis}}{{Schlegel} et~al.}{1998}]{schlegel1998}
{Schlegel} D.~J.,  {Finkbeiner} D.~P.,   {Davis} M.,  1998, \mn@doi [\apj]
  {10.1086/305772}, \href {http://adsabs.harvard.edu/abs/1998ApJ...500..525S}
  {500, 525}

\bibitem[\protect\citeauthoryear{{Spolaor}, {Proctor}, {Forbes}  \&
  {Couch}}{{Spolaor} et~al.}{2009}]{spolaor2009}
{Spolaor} M.,  {Proctor} R.~N.,  {Forbes} D.~A.,   {Couch} W.~J.,  2009,
  \mn@doi [\apjl] {10.1088/0004-637X/691/2/L138}, \href
  {http://adsabs.harvard.edu/abs/2009ApJ...691L.138S} {691, L138}

\bibitem[\protect\citeauthoryear{{Storchi-Bergmann}, {Riffel}, {Riffel},
  {Diniz}, {Borges Vale}  \& {McGregor}}{{Storchi-Bergmann}
  et~al.}{2012}]{brasil3}
{Storchi-Bergmann} T.,  {Riffel} R.~A.,  {Riffel} R.,  {Diniz} M.~R.,  {Borges
  Vale} T.,   {McGregor} P.~J.,  2012, \mn@doi [\apj]
  {10.1088/0004-637X/755/2/87}, \href
  {http://adsabs.harvard.edu/abs/2012ApJ...755...87S} {755, 87}

\bibitem[\protect\citeauthoryear{{Thomas}, {Maraston}, {Schawinski}, {Sarzi}
  \& {Silk}}{{Thomas} et~al.}{2010}]{thomas2010}
{Thomas} D.,  {Maraston} C.,  {Schawinski} K.,  {Sarzi} M.,   {Silk} J.,  2010,
  \mn@doi [\mnras] {10.1111/j.1365-2966.2010.16427.x}, \href
  {http://adsabs.harvard.edu/abs/2010MNRAS.404.1775T} {404, 1775}

\bibitem[\protect\citeauthoryear{{Tortora} \& {Napolitano}}{{Tortora} \&
  {Napolitano}}{2012}]{tortora2012}
{Tortora} C.,  {Napolitano} N.~R.,  2012, \mn@doi [\mnras]
  {10.1111/j.1365-2966.2012.20478.x}, \href
  {http://adsabs.harvard.edu/abs/2012MNRAS.421.2478T} {421, 2478}

\bibitem[\protect\citeauthoryear{{Wang}, {Mo}, {Jing}, {Guo}, {van den Bosch}
  \& {Yang}}{{Wang} et~al.}{2009}]{wang2009}
{Wang} H.,  {Mo} H.~J.,  {Jing} Y.~P.,  {Guo} Y.,  {van den Bosch} F.~C.,
  {Yang} X.,  2009, \mn@doi [\mnras] {10.1111/j.1365-2966.2008.14301.x}, \href
  {http://adsabs.harvard.edu/abs/2009MNRAS.394..398W} {394, 398}

\bibitem[\protect\citeauthoryear{{Wang}, {Mo}, {Yang}  \& {van den
  Bosch}}{{Wang} et~al.}{2012}]{wang2012}
{Wang} H.,  {Mo} H.~J.,  {Yang} X.,   {van den Bosch} F.~C.,  2012, \mn@doi
  [\mnras] {10.1111/j.1365-2966.2011.20174.x}, \href
  {http://adsabs.harvard.edu/abs/2012MNRAS.420.1809W} {420, 1809}

\bibitem[\protect\citeauthoryear{{White} \& {Rees}}{{White} \&
  {Rees}}{1978}]{white1978}
{White} S.~D.~M.,  {Rees} M.~J.,  1978, \mnras, \href
  {http://adsabs.harvard.edu/abs/1978MNRAS.183..341W} {183, 341}

\bibitem[\protect\citeauthoryear{{Wilkinson} et~al.,}{{Wilkinson}
  et~al.}{2015}]{wilkinson2015}
{Wilkinson} D.~M.,  et~al., 2015, \mn@doi [\mnras] {10.1093/mnras/stv301},
  \href {http://adsabs.harvard.edu/abs/2015MNRAS.449..328W} {449, 328}

\bibitem[\protect\citeauthoryear{{Yang}, {Mo}, {van den Bosch}, {Pasquali},
  {Li}  \& {Barden}}{{Yang} et~al.}{2007}]{yang2007}
{Yang} X.,  {Mo} H.~J.,  {van den Bosch} F.~C.,  {Pasquali} A.,  {Li} C.,
  {Barden} M.,  2007, \mn@doi [\apj] {10.1086/522027}, \href
  {http://adsabs.harvard.edu/abs/2007ApJ...671..153Y} {671, 153}

\bibitem[\protect\citeauthoryear{{York} et~al.,}{{York}
  et~al.}{2000}]{york2000}
{York} D.~G.,  et~al., 2000, \mn@doi [\aj] {10.1086/301513}, \href
  {http://adsabs.harvard.edu/abs/2000AJ....120.1579Y} {120, 1579}

\bibitem[\protect\citeauthoryear{{Zheng} et~al.}{{Zheng}
  et~al.}{2016}]{zheng2016}
{Zheng} Z.,  et~al., 2016, \apj, submitted

\bibitem[\protect\citeauthoryear{{de Zeeuw} et~al.,}{{de Zeeuw}
  et~al.}{2002}]{deze2002}
{de Zeeuw} P.~T.,  et~al., 2002, \mn@doi [\mnras]
  {10.1046/j.1365-8711.2002.05059.x}, \href
  {http://adsabs.harvard.edu/abs/2002MNRAS.329..513D} {329, 513}

\makeatother
\end{thebibliography}

\bsp	
\label{lastpage}
\end{document}